\documentclass[
amsmath, amssymb, 
aps, pra,
reprint, twocolumn,
superscriptaddress,
longbibliography,
floatfix,
showkeys
]{revtex4-1}

\usepackage{graphicx}
\usepackage{csquotes}
\usepackage[urlcolor=blue, colorlinks=true,citecolor=blue]{hyperref}

\usepackage{color}
\usepackage{bbm}
\usepackage{nicefrac}

\usepackage{upgreek}

\def\avg#1{\mathinner{\langle{#1}\rangle}}
\def\bra#1{\ensuremath{\mathinner{\langle{#1}|}}}
\def\ket#1{\ensuremath{\mathinner{|{#1}\rangle}}}

\newcommand{\expval}[1]{\langle #1\rangle}

\begin{document}
\title{Quantum chemistry calculations on a trapped-ion quantum simulator}

\author{Cornelius Hempel}
	\thanks{These authors contributed equally to this work.}
	\affiliation{Institut f\"ur Quantenoptik und Quanteninformation,
	\"Osterreichische Akademie der Wissenschaften, Technikerstr. 21A, 6020 Innsbruck,
	Austria}
	\affiliation{
	ARC Centre of Excellence for Engineered Quantum Systems, School of Physics, University of Sydney, NSW, 2006, Australia }
	\email{cornelius.hempel@gmail.com}
\author{Christine Maier}
	\thanks{These authors contributed equally to this work.}
	\affiliation{Institut f\"ur Quantenoptik und Quanteninformation,
	\"Osterreichische Akademie der Wissenschaften, Technikerstr. 21A, 6020 Innsbruck,
	Austria}
	\affiliation{
	Institut f\"ur Experimentalphysik, Universit\"at Innsbruck,	Technikerstr. 25, 6020 Innsbruck, Austria}
\author{Jonathan Romero}
	\affiliation{Department of Chemistry and Chemical Biology, Harvard University
	12 Oxford St, Cambridge, MA 02138, USA}
    \author{Jarrod McClean}
	\affiliation{Google Inc., 340 Main Street, Venice, CA 90291, USA}
\author{Thomas Monz}
	\affiliation{
	Institut f\"ur Experimentalphysik, Universit\"at Innsbruck,	Technikerstr. 25, 6020 Innsbruck, Austria}
\author{Heng Shen}
	\affiliation{Institut f\"ur Quantenoptik und Quanteninformation,
	\"Osterreichische Akademie der Wissenschaften, Technikerstr. 21A, 6020 Innsbruck,
	Austria}
	\affiliation{
	Institut f\"ur Experimentalphysik, Universit\"at Innsbruck,	Technikerstr. 25, 6020 Innsbruck, Austria}
\author{Petar Jurcevic}
	\affiliation{Institut f\"ur Quantenoptik und Quanteninformation,
	\"Osterreichische Akademie der Wissenschaften, Technikerstr. 21A, 6020 Innsbruck,
	Austria}
	\affiliation{
	Institut f\"ur Experimentalphysik, Universit\"at Innsbruck,	Technikerstr. 25, 6020 Innsbruck, Austria}
 \author{Ben P. Lanyon}
 	\affiliation{Institut f\"ur Quantenoptik und Quanteninformation,
 	\"Osterreichische Akademie der Wissenschaften, Technikerstr. 21A, 6020 Innsbruck,
 	Austria}
	\affiliation{
	Institut f\"ur Experimentalphysik, Universit\"at Innsbruck,	Technikerstr. 25, 6020 Innsbruck, Austria}
\author{Peter Love}
	\affiliation{Department of Physics and Astronomy, Tufts University, 574 Boston Ave., Medford, MA 02155, USA}
                \author{Ryan Babbush}
	\affiliation{Google Inc., 340 Main Street, Venice, CA 90291, USA}
\author{Al\'{a}n Aspuru-Guzik}
	\affiliation{Department of Chemistry and Chemical Biology, Harvard University
	12 Oxford St, Cambridge, MA 02138, USA}
\author{Rainer Blatt}
	\affiliation{Institut f\"ur Quantenoptik und Quanteninformation,
	\"Osterreichische Akademie der Wissenschaften, Technikerstr. 21A, 6020 Innsbruck,
	Austria}
	\affiliation{
	Institut f\"ur Experimentalphysik, Universit\"at Innsbruck,	Technikerstr. 25, 6020 Innsbruck, Austria}
\author{Christian F. Roos}
	\affiliation{Institut f\"ur Quantenoptik und Quanteninformation,
	\"Osterreichische Akademie der Wissenschaften, Technikerstr. 21A, 6020 Innsbruck,
	Austria}
	\affiliation{
	Institut f\"ur Experimentalphysik, Universit\"at Innsbruck,	Technikerstr. 25, 6020 Innsbruck, Austria}
	\email{christian.roos@uibk.ac.at}
\date{\today}     

\begin{abstract}
Quantum-classical hybrid algorithms are emerging as promising candidates for near-term practical applications of quantum information processors in a wide variety of fields ranging from chemistry to physics and materials science. 
We report on the experimental implementation of such an algorithm to solve a quantum chemistry problem, using a digital quantum simulator based on trapped ions. 
Specifically, we implement the variational quantum eigensolver algorithm to calculate the molecular ground state energies of two simple molecules and experimentally demonstrate and compare different encoding methods using up to four qubits.
Furthermore, we discuss the impact of measurement noise as well as mitigation strategies and indicate the potential for adaptive implementations focused on reaching chemical accuracy, which may serve as a cross-platform benchmark for multi-qubit quantum simulators.
\end{abstract}

\maketitle

\section{Introduction}
Quantum simulators \cite{Buluta:2009, naturephys, Schaetz:2013, Georgescu:2014, Johnson:2014} are widely recognized for their promise to help solve challenging problems in a range of fields, including physics, chemistry and materials science. 
First suggested by Richard Feynman in the early 1980s~\cite{Feynman:1982}, quantum simulators aim to harness controlled quantum evolutions in order to simulate other quantum systems. Later refined to a formalized notion of a universal quantum computer~\cite{Lloyd:1996}, a quantum simulator's main purpose is to turn the exponential scaling of resources needed to simulate quantum systems on classical computers into a more favorable polynomial overhead on a quantum machine.
Experimentally pioneered at the turn of the millennium by nuclear magnetic resonance~\cite{Somaroo:1999}, neutral atom~\cite{Greiner:2002} and trapped-ion~\cite{Leibfried:2002} systems, quantum simulation has attracted much attention over the last decade and might well hold the key to unlock significant advances in our understanding of many-body physics~\cite{CiracZoller:2012}. 

In the analog approach to quantum simulation, the interactions of a quantum system are engineered to closely match the Hamiltonian of the system of interest. Starting from a well-defined initial state, the simulator evolves under a tailored Hamiltonian to a final state, and measurements are taken that reveal the sought-after information about the system of interest. Prominent examples  are experiments realizing Bose- or Fermi-Hubbard models with ultracold atoms~\cite{Gross:2017}, or long-range spin models using either up to 20 fully controlled~\cite{Islam:2013,Jurcevic:2014,Richerme:2014,Jurcevic:2015,Smith:2016,Friis:2017} or over 100 entangled ions~\cite{Britton:2012, Bohnet:2016, Zhang:2017}. 

The digital approach to quantum simulation employs computational gates to approximate the time evolution of arbitrary local Hamiltonians. 
This can be done in a freely programmable way~\cite{Lloyd:1996} that is amenable to quantum error correction \cite{QEC:2013,Terhal:2015}. With a versatility similar to classical computers, digital quantum simulators offer the prospect to shed light on properties of systems that are fundamentally different from the simulator. Laser-cooled ensembles of trapped ions that simulate high-energy physics processes are an example~\cite{Martinez:2016b}. The first multi-qubit demonstrations of the technique on scalable platforms were implemented on trapped ions~\cite{Lanyon:2011} and a superconducting system \cite{Barends:2015}.

To enable quantum simulations with a level of complexity that is beyond the reach of classical computers, multiple avenues are being pursued to scale up universal quantum processors realized on platforms such as trapped ions or superconducting qubits~\cite{Monroe:2013, Devoret:2013}. As experimental progress paves the way towards the realization of larger scale devices, it is important to develop algorithms that make efficient use of these continually improving quantum resources. This is where quantum-classical hybrid algorithms have recently emerged, which leverage the unique capabilities of quantum devices by incorporating them in classical numerical calculations. 

In this article, we use a digital quantum simulator based on trapped ions to experimentally investigate one such algorithm,  the variational quantum eigensolver~\mbox{\cite{Yung:2014, Peruzzo:2014, McClean:2016}} for the calculation of molecular ground state energies. We begin with a review of the underlying methods of quantum chemistry as well as the overall algorithmic approach and briefly summarize previous experimental implementations. We then introduce our experimental realization and present results for two example molecules, molecular hydrogen and lithium hydride. Finally, we discuss the results and sources of error with respect to a threshold known as chemical accuracy and provide suggestions for further improvements of the algorithm. 

\section{Simulating Quantum Chemistry}

\subsection{Approach and specific steps}

While quantum computers have been shown to offer exponential speedups for a variety of problems, a particularly compelling application is the quantum computation of molecular energies \cite{Aspuru-Guzik:2005,Abrams:1997,Abrams:1999}.  Efficient quantum simulations of classically intractable instances of the associated electronic structure problem promise breakthroughs in our understanding of basic chemistry and could revolutionize research into new materials, pharmaceuticals, and industrial catalysts. In the last few years, many efforts have sought to develop new algorithms \cite{Kassal:2008,Whitfield:2010,Babbush:2014,Babbush:2016,Kivlichan:2016,Babbush:2017,Kivlichan:2018} and better implementation strategies \cite{Jones:2012,McClean:2014,Wecker:2014,Hastings:2015,Poulin:2015, Babbush:2015,Reiher:2017,Babbush:2018} for these simulations. Among the explored approaches, the variational quantum eigensolver (VQE) algorithm \cite{Yung:2014, Peruzzo:2014, McClean:2016} has been shown experimentally to be resilient to some systematic errors~\cite{OMalley:2016} from imperfect control pulses, making it a promising candidate for quantum simulations in the near-term~\cite{Wecker:2015}.

VQEs belong to the class of quantum-classical hybrid algorithms where a classical subroutine is enhanced by the computational power of a quantum simulator. Given its potential for near-term use, the VQE approach is increasingly receiving attention in both theoretical \cite{Barrett:2013, Bauer:2016, Wecker:2015,Kreula:2016, McClean:2016, McClean:2017,Li:2017, Romero:2017,Babbush:2017,Wang:2018,Rubin:2018} and experimental works \cite{Peruzzo:2014, Eichler:2015, Wang:2015, Shen:2017, OMalley:2016, Kandala:2017, Colless:2017, Dumitrescu:2018} with a focus on both quantum chemistry as well as fundamental physics and materials science. We now discuss the simulation of quantum chemistry in more detail.

The central problem of theoretical chemistry is to compute the lowest energy eigenvalue of the molecular electronic structure Hamiltonian. The eigenstates of this Hamiltonian determine almost all of the properties of interest in a molecule or material, and as the energy gap between the ground and first excited state is often much larger than 25.7~meV ($\mathrm{k_B} T$ at room temperature), the ground state is of particular interest. To arrive at the standard form of this Hamiltonian used in quantum computation, one begins from a collection of nuclear charges $Z_i$ and a number of electrons in the system for which the corresponding Hamiltonian is written as
\begin{equation}
\label{eq:firstQHam}
\begin{aligned}
H_1 &= - \sum_i \frac{\nabla_{R_i}^2 }{2M_i} - \sum_i \frac{\nabla_{r_i}^2}{2} - \sum_{i,j} \frac{Z_i}{|R_i - r_j|} \\
&+ \sum_{i, j > i} \frac{Z_i Z_j}{|R_i - R_j|} + \sum_{i, j>i} \frac{1}{|r_i - r_j|}
\end{aligned} 
\end{equation}
in atomic units $(\hbar = 1)$. Here the positions, masses, and charges of the nuclei are $R_i, M_i, Z_i$, and the positions of the electrons are $r_i$. This form of the Hamiltonian and its real-space discretization is often referred to as the first-quantized formulation of quantum chemistry and generally enforces the fermionic nature of the electron through an anti-symmetric wavefunction.
Several approaches have been developed for treating this form of the problem on a quantum computer~\cite{Kassal:2008,Kassal:2010,Toloui:2013,Bravyi:2017,BabbushSymmetry:2017,BabbushSparse2:2018}; however, the focus of this work is on the second-quantized formulation.

To reach the second quantized formulation, one typically first approximates the nuclei as fixed classical point charges under the Born-Oppenheimer approximation and chooses a basis $\phi_i$ in which to represent the electronic wavefunction. Often, one chooses a basis of $N$ molecular orbitals, constructed as a linear combination of atomic orbitals (LCAO), which are computed using a mean-field procedure known in chemistry as the Hartree-Fock method~\cite{Helgaker:2000}. The atomic orbital basis functions are derived from variations of hydrogen-like atomic orbitals for different values of $Z$. They are numerically optimized to match desired physical properties across a range of systems and to be compatible with systematic improvement~\cite{Kendall:1992}. Usually the basis functions are expressed as sums of Gaussian functions rather than the original Slater type orbitals to enhance the efficiency of integral evaluation. This choice is convenient for small problems, but not mandatory and much work has been done recently to improve the Hamiltonian representation for electronic structure problems~\cite{Babbush:2017,Kivlichan:2018}. 

The second-quantized formulation of quantum chemistry leads to the (electronic) Hamiltonian  
\begin{align}
H_2 = \sum_{pq} h_{pq} a^{\dagger}_p a_q + \frac{1}{2} \sum_{pqrs} h_{pqrs} a^{\dagger}_p a^{\dagger}_q a_r a_s, \label{eq:ferm_hamiltonian}
\end{align}
in which the wavefunction's anti-symmetry is enforced through the anti-commutation relations of the fermion creation and annihilation operators $a_i^\dagger$ and $a_j$.

The goal now is to compute the energy of electrons interacting in the fixed external potential of the atomic nuclei. Encoding the electron's spatial and spin coordinate as \mbox{$\sigma_i = (r_i, s_i)$}, the two sets of scalar coefficients in Eq.~\eqref{eq:ferm_hamiltonian} can be calculated via  
\begin{equation}
\label{eq:integrals}
\begin{aligned}
h_{pq} &= \int \ d\sigma \ \phi_p^*(\sigma) \left(\frac{\nabla_r^2}{2} - \sum_i \frac{Z_i}{|R_i - r|} \right)\phi_q^*(\sigma)  \\
h_{pqrs} &= \int \ d\sigma_1 \ d\sigma_2 \frac{\phi_p^*(\sigma_1)\phi_q^*(\sigma_2) \phi_s(\sigma_1)\phi_r(\sigma_2)}{|r_1 - r_2|}. \end{aligned}
\end{equation}

In order to implement the Hamiltonian on a qubit-based quantum simulator, the specific fermionic Hamiltonians need to be transformed to spin Hamiltonians. The most common schemes for this transformation are the Jordan-Wigner transformation~\cite{Jordan:1928,Somma:2002} and the Bravyi-Kitaev transformation~\cite{Somma:2002,Bravyi:2002,Seeley:2012,Tranter:2015,Havlicek:2017,Bravyi:2017,Steudtner:2017}. Characteristically, the Jordan-Wigner (JW) transformation leads to $N$-local Hamiltonians and the Bravyi-Kitaev (BK) transformation leads to $\log (N)$-local Hamiltonians. In this work, we explore both approaches to arrive at a spin Hamiltonian $H$, which can be implemented with qubits on a quantum simulator.

\begin{figure}[!t] 
   \centering
   \includegraphics[scale=1]{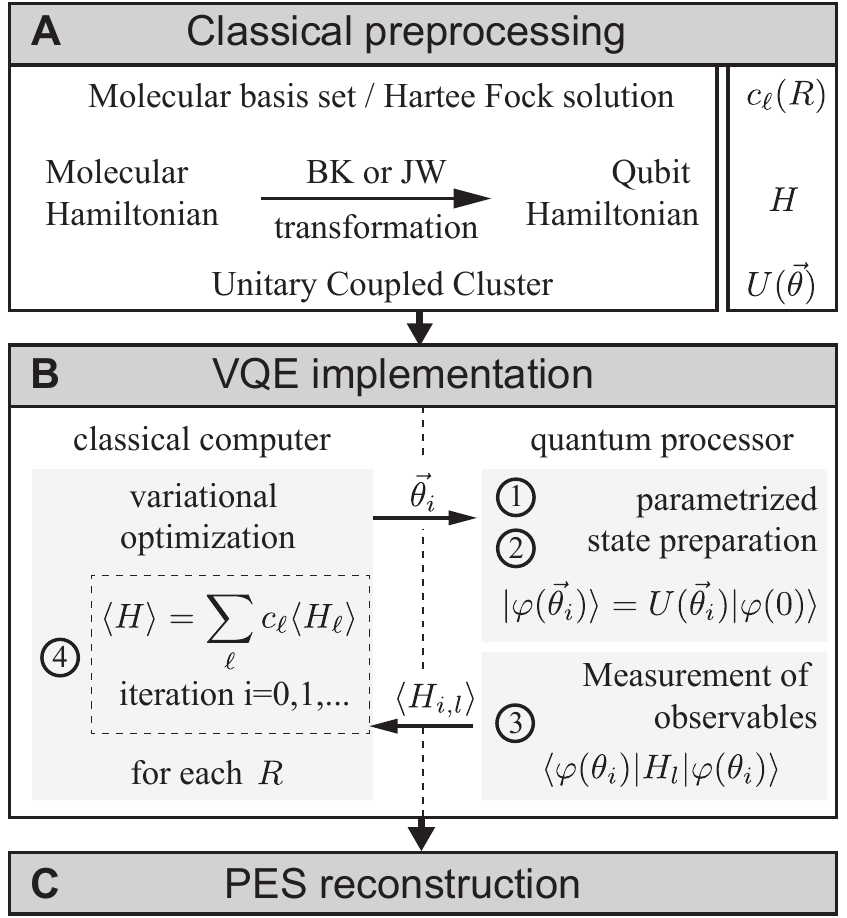} 
   \caption{Steps of a VQE-based quantum chemistry calculation. A classical preprocessing stage \textbf{(A)} translates the chemical problem to a classical description of a quantum circuit for unitary operator $U(\vec{\theta})$ that can be directly implemented on a quantum processor. \textbf{(B)} As outlined in the main text above, the four steps of the iterative variational optimization are performed in a loop combining quantum resources with a numerical search running on a classical computer. Finally, in \textbf{(C)}, the full potential energy surface (PES) of the molecule in question, including both nuclear and electronic contributions, is reconstructed from the data.}
   \label{fig:VQE}
\end{figure}

VQE algorithms for quantum chemistry typically seek to prepare the ground state of the target system for a particular geometric configuration specified by $\vec{R_i}$. The energy landscape created by this geometry for the electrons, derives from the integrals in Eqs.~\eqref{eq:integrals} above and will be captured by scalar values $c_\ell$ throughout the remainder of the article. The calculation then proceeds along the following four steps, visually summarized in Fig.~\ref{fig:VQE}.

\begin{enumerate}
\item A digital quantum simulator is initialized in a simple state $\ket{\varphi(0)}$ which represents a good classical approximation to the ground state of Hamiltonian $H$. When using a molecular orbital basis, this initialization is particularly straightforward as the classical mean-field state (most often a Hartree-Fock solution~\cite{Helgaker:2000}) is a product state. 
\item A quantum circuit implementing the unitary operation $U(\vec \theta_0)$ is applied to $\ket{\varphi(0)}$ mapping the initial state to a parameterized ``ansatz'' state \mbox{$\ket{\varphi(\vec \theta_0)} = U(\vec \theta_0) \ket{\varphi(0)}$}. The operation $U(\vec \theta_0)$ and its parameter vector $\vec \theta_0$ are chosen based on known structure in the target system, which is usually obtained from classical approximation methods. 
\item One measures the expectation value of the energy $\avg{H} = \bra{\varphi(\vec \theta_0)} H \ket{\varphi(\vec \theta_0)}$ of the prepared ansatz state. This makes use of the form of the Hamiltonian $H = \sum_{\ell} c_\ell H_\ell$, where $H_\ell$ are tensor products of Pauli matrices and $c_\ell$ the above mentioned scalars that were pre-calculated for a given internuclear configuration $\vec{R_i}$. Repeated rounds of state preparation and measurement of the individual terms $\avg{H_\ell}$ allow one to provide an estimate for the expectation value $\avg{H} = \sum_\ell c_\ell \avg{H_\ell}$. In this context, this step is often referred to as Hamiltonian averaging~\cite{Peruzzo:2014,McClean:2016}.
\item One adjusts the parameters $\vec \theta$ to minimize $\avg{H}$. An iterative classical ``outer-loop'' optimization (e.g. gradient descent) can be deployed for this purpose. Assuming this procedure converges after $m$ iterations, the resulting state $\ket{\varphi(\vec \theta_m)}$ represents the variational approximation to the molecular ground state at the chosen configuration $\vec{R_i}$.
\end{enumerate}

It is worth noting that, when applied to classical (diagonal) Hamiltonians, the above strategy is the basis of the quantum approximate optimization algorithm \cite{Farhi:2014}. 

In order to obtain the total molecular energy of the potential energy surface, the mutual Coulomb energy of the nuclei, separated out in the Born-Oppenheimer approximation of the classical preprocessing step before, has to be added to the VQE result at every configuration $\vec{R_i}$.

A good ansatz unitary $U(\vec \theta)$ is typically chosen to balance the strengths of available experimental resources with insights about the structure of the fermionic system under study. For instance, it has been argued that a parameterized Trotter approximation~\cite{Trotter:1959, Lloyd:1996}, to the adiabatic state preparation algorithm may serve as an effective ansatz \cite{Wecker:2015,Babbush:2017,Kivlichan:2018}. 
In this article, we parameterize our variational ansatz based on insights from a cornerstone method of modern electronic structure theory known as coupled cluster~\cite{Helgaker:2002}. While widely regarded the ``gold standard'' of accuracy for the classical study of strongly correlated systems, a drawback of the coupled cluster ansatz is that the method is non-unitary and there is no way to ensure that solutions are properly normalized. 
This can lead to significant errors when the ground state has what is referred to as multireference character, i.e. when it has significant support on more than one classical configuration within the chosen basis. To address this problem, a unitary variant of the coupled cluster method was proposed in \cite{Bartlett:1989}. However, while theoretically immune to some of the problems of the traditional method, unitary coupled cluster (UCC) calculations cannot be performed efficiently on classical computers. Thus, the classical study of UCC involves additional approximations which are also problematic in some cases~\cite{Taube:2006}. However, as first discussed in \cite{Peruzzo:2014}, one can perform finite order UCC calculations efficiently on a quantum computer without any approximation \cite{McClean:2016, Romero:2017}.

UCC essentially suggests a series of fermionic operators that one should evolve under in order to prepare the ground state. For instance, one can think of the unitary coupled cluster ansatz in a Trotter approximation as
\begin{equation}
U_\mathrm{UCC}\left(\vec \theta\right) = \mathrm{e}^{\left\{\sum_\gamma \theta_\gamma ( T_\gamma - T_\gamma^\dagger)\right\}}\approx \prod_{\gamma} \mathrm{e}^{\theta_\gamma( T_\gamma - T_\gamma^\dagger)}, 
\label{eq:UCCferm}
\end{equation}
where $\left(T_\gamma - T_\gamma^\dagger\right)$ are anti-Hermitian fermionic excitation operators. The approximation associated with using only a single Trotter step for the exponential is generally acceptable since the coupled cluster amplitudes $\theta_\gamma$ are usually quite small. As discussed in \cite{Romero:2017}, one can scalably deploy classical coupled cluster calculations in order to get a reasonable initial guess of the amplitudes $\theta_0$. Furthermore, due to selection rules involving the symmetry point groups of molecular orbitals and suitable quantum numbers of the Hamiltonian, most of the coupled cluster amplitudes associated with operators that one might naively anticipate will need to be simulated, actually are zero. A vast majority of the other amplitudes are typically extremely small~\cite{Romero:2017}, such that one may choose a threshold value for the $\vec \theta_0$ provided by the classical coupled cluster method to further reduce the number of relevant $e^{\theta_\gamma ( T_\gamma - T_\gamma^\dagger)}$ terms.

\subsection{Previous demonstrations}

Experimental demonstrations of quantum chemistry algorithms have been performed on architectures ranging from quantum photonics~\cite{Lanyon:2010, Peruzzo:2014, Paesani:2017}, NV centers \cite{Wang:2015}, ion traps \cite{Shen:2017, Shen:2018} and NMR platforms \cite{Du:2010,Li:2011} to superconducting qubits~\cite{OMalley:2016, Kandala:2017, Colless:2017}.
The first realizations date back to 2010, when a photonic experiment~\cite{Lanyon:2010} and a closely related NMR experiment~\cite{Du:2010} simulated the hydrogen molecule in the simplest basis, using a method that required considerable classical precomputation. A similar experiment computing the dissociation curve of the HeH$^+$ cation was performed using a nitrogen vacancy based system in 2015~\cite{Wang:2015} and the first VQE experiment investigated the same molecule in a minimal basis in 2013~\cite{Peruzzo:2014} using a photonic implementation. 

The first scalable quantum chemistry simulation was carried out on a superconducting platform in 2016~\cite{OMalley:2016}. Here scalable means that the required classical precomputation resources increase  in the same way that they would for an arbitrarily large problem. 
The work in \cite{OMalley:2016} used a one-parameter variational ansatz and performed VQE in post-processing to calculate the ground state energy of H$_2$, reaching chemical accuracy for the binding energy (deviations $\leq1.6\times 10^{-3}$~Hartree). The authors also implemented the phase estimation algorithm for quantum chemistry \cite{Aspuru-Guzik:2005}, which did not reach chemical accuracy due to the degradation of gate fidelity in the absence of error correction on their device. As both experiments were performed on the same chip using gates of the same fidelity, these results provide a first validation that the VQE approaches we focus on can deliver superior results on current hardware.

The most advanced implementation of the VQE algorithm in terms of molecular size and resource scaling was published 2017 in the work reported in~\cite{Kandala:2017}, which simulated three molecules H$_2$, LiH and BeH$_2$ on a superconducting qubit platform. The authors also performed a theoretical investigation of the resource requirements in terms of circuit depth for each molecule and found a significantly more favorable scaling for devices with all-to-all qubit connectivity, validating the potential utility of ion trap hardware with its native all-to-all connectivity for these types of simulations. Most recently a superconducting implementation of a modified VQE algorithm was used to also determine excited states of the H$_2$ molecule~\cite{Colless:2017}.

Ion-trap implementations of quantum simulation applied to quantum chemistry have so far been limited to working with a single qubit. The work in reported in~\cite{Shen:2017} used a single $^{171}{\rm Yb}^+$ ion and mapped the Hamiltonian of the molecule to four of its internal energy levels, which is not a scalable procedure and provides no advantage over classical computation~\cite{Meyer:2000}. However, the other steps of the VQE algorithm were performed scalably, i.e. as they would be for a different state mapping. Recently, the same system was used to explore vibronic molecular spectra \cite{Shen:2018} employing the bosonic motional degree of freedom of a single trapped ion.

VQE is not the only method that reduces the experimental overhead of phase estimation. Recently, Bayesian approaches to phase estimation have been proposed~\cite{wiebe2016Bayesian}. This method has been implemented in a photonic simulation of minimal basis H$_2$ system - reaching chemical accuracy~\cite{Paesani:2017}, however with scalability properties comparable to~\cite{Lanyon:2010}. 

\section{Experimental Implementation}
The trapped ion system used to implement our digital quantum simulation consists of a linear Paul trap in which a variable number of $^{40}$Ca$^+$ ions are electrically confined for many days at a time.  Each ion stores a quantum bit that is encoded in a pair of Zeeman states chosen from their $4\mathrm{S}_{1/2}$ electronic ground and $3\mathrm{D}_{5/2}$ metastable states. In the following, these electronic states will be labeled as qubit states $\ket{1}$ and $\ket{0}$, respectively.

\begin{figure*}[t!] 
   \centering
   \includegraphics[scale=1]{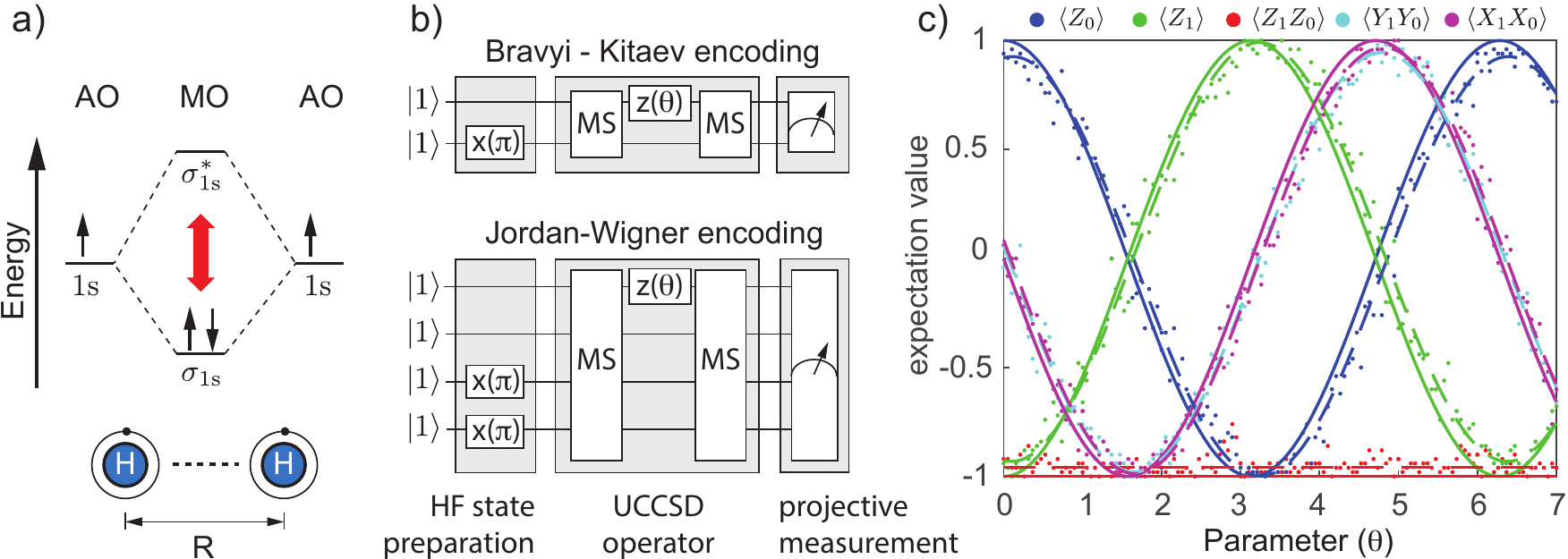} 
   \caption{\textbf{a)} Molecular orbitals (MOs) of the hydrogen molecule, built from the individual atoms' 1s atomic orbitals (AOs). The electronic energy is calculated for nuclei separated by a distance $R$. 
   \textbf{b)} Quantum circuits implementing the $U_\textrm{UCCSD}$ operator with respect to a Hartree-Fock (HF) ansatz state for two different mappings from fermions to spins.
     \textbf{c)} Expectation values obtained in the experiment under the two-qubit Bravyi-Kitaev encoding as a function of unitary circuit parameter $\theta$. Solid lines correspond to theoretical predictions, while dashed lines are fits to the experimental data (dots) obtained with 100 repetitions per point. Error bars from quantum projection noise are omitted for clarity but taken into account in the fitting routine.}
   \label{fig:H2-orb-circ-expval}
\end{figure*}

The qubits are manipulated via a set of global and tightly-focussed, addressed laser beams. Single qubit gate operations are implemented via a three-pulse-sequence that combines two global qubit rotations with an intermediate addressed laser pulse manipulating only the targeted qubit~\cite{Schindler:2013}. Multi-qubit entangling operations are realized through laser-driven interactions that are mediated by the collective motional modes of the ions within their common trapping potential. 

After mapping the UCC excitation operators in Eq.~\eqref{eq:UCCferm} to tensor products of Pauli operators, we implement each operator using appropriate quantum circuits, specifically illustrated for each case below. The circuits' general structure combines two multi-qubit entangling gates, each realizing the unitary \mbox{$U_\mathrm{MS}(\phi)=\exp(-i\phi/2\sum_{i<j}\sigma^i_x\sigma^j_x)$}, with a single qubit operation that encodes a single element of the parameter vector $\vec{\theta}$ in Fig.~\ref{fig:VQE} and a particular $\theta_\gamma$ in Eq.~\eqref{eq:UCCferm}, respectively. The subscript MS stands for M\o{}lmer and S\o{}rensen, the originators~\cite{Sorensen:1999, Sorensen:2000} of this type of entangling gate and $\phi=\pi/2$ for the creation of an entangled state, when starting in any state of the computational basis ($z$). As a unit, this circuit building block~\cite{Muller:2011, Casanova:2012} can effectively realize arbitrary many-body interactions as required by the Pauli operators $H_\ell$ resulting from the transformed UCC operators. 

The estimation of the expectation values of individual Pauli operators $\langle H_\ell \rangle$ is carried out through a projective measurement in the logical $z$-basis. This measurement is implemented  using laser-induced, state-dependent fluorescence which we detect on a CCD camera, allowing for simultaneous readout of all qubits (cf. Appendix~\ref{ap:qubit_implementation}). To determine $\langle H_\ell \rangle$ with terms in bases other than $z$, we employ suitable global and single qubit operations to rotate these into the measurement basis prior to the projective measurement. We now discuss the individual steps and results for the two example molecules investigated experimentally.

\section{Molecular hydrogen}

The simplest possible neutral molecule is formed by two hydrogen atoms. We aim to calculate the potential energy surface of its ground state as function of the internuclear distance $R$. This will be done in two ways, (1) by scanning the entire parameter space associated with the UCC operator of the spin-transformed molecular Hamiltonian $H$ and (2) by running the VQE algorithm for different values of $R$ separately,  leading to only a sparse sampling of this parameter space. Due to its simplicity, H$_2$ has become a standard example for experimental implementations of quantum chemistry algorithms. 

\subsection{Encoding the problem}

A common molecular basis set (Figure~\ref{fig:VQE}, step A), is that of Slater-type orbitals (STO), represented by linear combinations of, e.g., three Gaussian functions (STO-3G). In this minimal basis set, each hydrogen atom contributes a 1s atomic orbital. Molecular orbitals are then formed by adding the corresponding two atomic wavefunctions. In-phase addition results in a lower energy $\sigma_{1s}$ bonding orbital with increased electron density between the nuclei and out-of-phase addition in a $\sigma^*_{1s}$ \mbox{anti-bonding} orbital associated with a depletion in electron density between them (Figure~\ref{fig:H2-orb-circ-expval}.a). 

Using this minimal basis set, a classical Hartree-Fock calculation allows us to pre-compute the relevant molecular orbitals for each geometric configuration specified by the respective internuclear separation $R$. This numerical step determines the scalars $c_\ell(R)$ by solving the integrals in Eq.~\eqref{eq:integrals} capturing the spatial and spin coordinates. It also yields a product state solution for the molecular wavefunction, which is used as an initial guess $\ket{\varphi(0)}$ in Figure~\ref{fig:VQE}'s step B. Following a series of theoretical considerations detailed in Appendix~\ref{ap:UCCH2}, we determine the only relevant unitary coupled cluster operator for single and double excitations (UCCSD), which can be expressed as,
\begin{equation}
\mbox{$U_\mathrm{UCCSD}(\theta)=e^{\theta (a_2^{\dag}a_3^{\dag}a_1a_0-a_0^{\dag}a_1^{\dag}a_3a_2)}.$}
\label{eq:H2double_exc_op}
\end{equation}
Here, $\theta$ is the parameter we aim to optimize, the indices correspond to the minimal set of orbitals and $a$ ($a^\dag$) to fermionic annihilation (creation) operators. Using the Jordan-Wigner transformation the fermionic operators of Eq.~\eqref{eq:H2double_exc_op} can be mapped to Pauli operators acting on four qubits as 
\begin{equation}
U^{\mathrm{JW}}_\mathrm{UCCSD}(\theta) = e^{-i\theta \sigma^x_3 \sigma^x_2\sigma^x_1\sigma^y_0} \nonumber
\end{equation}
with the initial Hartree-Fock state \mbox{$\ket{\varphi(0)_\mathrm{JW}} =\ket{0011}$}. Alternatively, after employing the Bravyi-Kitaev transformation, one can make use of the fact that after the mapping only qubits 0 and 2 are affected by operators other than $\sigma_z$ and $\mathcal{I}$, allowing the number of required qubits to be reduced to two \cite{OMalley:2016}. The resulting unitary   
\begin{equation}
U^\mathrm{BK}_\mathrm{UCCSD} (\theta) = e^{-i\theta \sigma_2^x\sigma_0^y} \nonumber
\end{equation}
 acts on the Hartree-Fock ansatz state $\ket{\varphi(0)_\mathrm{BK}} =\ket{01}$.

Each transformed UCC operator is implemented using the corresponding circuit shown in Figure~\ref{fig:H2-orb-circ-expval}.b. They only differ in the number of required qubits and the projective measurements prescribed by the Pauli operators $H_\ell$ in the respectively transformed Hamiltonians. Each circuit contains two single shot entangling gates \enquote{MS} that make use of the all-to-all connectivity of our device, making it particularly resource efficient for the larger register size of the four qubit implementation. Following steps outlined in Appendix~\ref{ap:UCCH2}, the effective Hamiltonian under the BK transformation becomes 
\begin{align}
H^\mathrm{BK}=&c_0 \mathcal{I} + c_1\sigma^{z}_{0}+c_2 \sigma^{z}_{1}\notag \\ &+ c_3 \sigma^{z}_{0}\sigma^{z}_{1}+ c_4 \sigma^{x}_{0}\sigma^{x}_{1} + c_5 \sigma^{y}_{0}\sigma^{y}_{1},
\label{eq:Ham_H2BK}
\end{align}
where the coefficients $c_\ell$ were all derived in this classical pre-processing step and $c_0$ specifically captures the spatially fixed nuclei's Coulomb potential. It correspondingly mandates three projective measurement settings to obtain the associated set of expectation values $H^\mathrm{BK}_\ell~=~\{Z_0, Z_1, X_0 X_1, Y_0 Y_1, Z_0 Z_1\}$. In the JW-transformed case, we are to obtain 14 expectation values {$H^\mathrm{JW}_\ell~=~\{Z_0, Z_1, Z_2, Z_3, Z_1Z_0, Z_2Z_0, Z_2Z_1, Z_3Z_0, Z_3Z_1,\\ Z_3Z_2, Y_3Y_2X_1X_0, Y_3X_2X_1Y_0, X_3Y_2Y_1X_0, X_3X_2Y_1Y_0\}$} from 5 different projective measurements. 

\begin{figure*}[!ht] 
   \centering
   \includegraphics[scale=1]{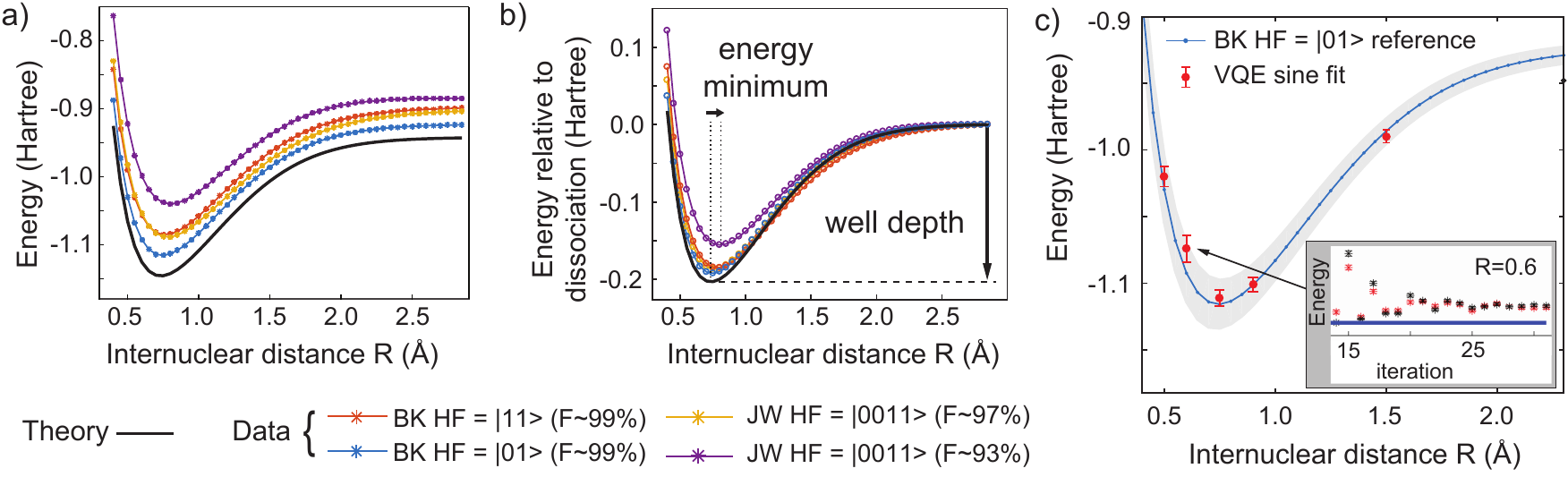} 
   \caption{\textbf{a)} Potential energy curves of the molecular hydrogen ground state. The black line corresponds to the theoretical value calculated in the chosen minimal basis. All other lines are derived from weighted sinusoidal fits to the energy surfaces formed from the experimentally obtained expectation values. The data sets vary the number of qubits, the Hartree Fock input states, encodings and gate fidelities as listed in the legend below the figures. \textbf{b)} Data from panel a) normalized to the theoretical dissociation energy at large internuclear separations $R$. The dashed and dotted lines indicate the well depth associated with the binding energy of the molecule and the position of the energy minimum, respectively. \textbf{c)} VQE implementation. The BK $ \rm{HF} = |01\rangle$ parameter scan fit result is shown as experimental reference with its 1$\sigma$ confidence band. Points with error bars indicate the five VQE runs performed. The inset shows the last iterations of one particular run that failed to converge to the target value (blue line) with experimental data depicted by red and a noise-free circuit simulation by black symbols.}
   \label{fig:H2-fringe-scan-results}
\end{figure*}

\subsection{Results}
\label{sec:ResultsH2}
With only a single parameter $\theta$ in both circuits, it is possible to efficiently scan the complete parameter space. Figure~\ref{fig:H2-orb-circ-expval}.c compares experimental results of such a parameter scan for the two qubit case under the BK transformation with theoretical predictions from a noise-free simulation of the circuit. In this way, measurements of a limited number of values for $\theta$ can be extrapolated to estimate the entire parameter range via Gaussian process regression \cite{OMalley:2016}, or by fitting sinusoidal functions to either the expectation values or the resulting energy landscape, as was done here. 
Once the expectation values $\expval{H_\ell(\theta)}$ are known over the entire parameter space spanned by $\theta$, one can calculate the molecular energies over the given range of internuclear distances $R$ 
yielding 
\begin{equation}
\expval{H(R,\theta)} = c_0(R) + \sum_\ell c_\ell(R) \expval{H_\ell(\theta)}.
\label{eq:energy}
\end{equation}

A ground state potential energy curve is then obtained by performing the energy minimization procedure in post-processing, effectively generating an experimental reference for the average performance of our simulator. The resulting energies are usually expressed in Hartree in chemistry, where $1~\text{Ha}=\nicefrac{\hbar^2}{m_e a_0} \approx 27.2~\text{eV}$, with Planck's constant $\hbar$, the mass of an electron $m_\text{e}$ and the Bohr radius $a_0$.

Under the influence of the MS gate, both the Hartree-Fock reference states $\ket{01}$ and $\ket{0011}$  are transformed through a decoherence-free subspace~\cite{Lidar:1998, Roos:2006, Lidar:2014} that is protected against correlated dephasing - a leading source of error in trapped ion qubit implementations. To experimentally investigate the impact of this decoherence channel on the results, we also implement a simulation that operates outside of the protected subspace. This is achieved by performing a basis rotation at the level of the Hamiltonian, which results in sign changes of the $c_\ell(R)$ coefficients and a different initial state. In the Bravyi-Kitaev case, this corresponds to simply changing the ansatz function to $\ket{\varphi_\mathrm{BK}(0)} = \ket{11}$, while still maintaining the same UCCSD operator and circuit implementation. More details on the construction of this particular decoherence free subspace (DFS) on our architecture can be found in reference \cite{Monz:2009}.

In total, we implement four cases in our demonstration, whose results are depicted in Figure~\ref{fig:H2-fringe-scan-results}.a and b on the next page. They include a DFS-protected and an unprotected implementation of the Bravyi-Kitaev mapping with a two-qubit MS gate fidelity of \mbox{$99(3)\%$} as well as a Jordan-Wigner implementation using a four-qubit MS gate with either $97(4)\%$ or $93(3)\%$ fidelity. In each case, the MS gate entanglement fidelity was estimated from population averages and the parity contrast relating to the coherence of the respective Bell or GHZ state generated from \ket{11} or \ket{1111} at the conclusion of the operation \cite{Sackett:2000, Leibfried:2003, Benhelm:2008}. 

We observe that the absolute energy values calculated from the parameter scans and illustrated in Figure~\ref{fig:H2-fringe-scan-results}.a are shifted to larger values both under increased correlated dephasing error as well as reduced gate fidelities. In particular, we observe that a two-qubit ansatz state that is not protected against correlated dephasing yields results similar to those of a DFS-protected four-qubit ansatz state. The latter is significantly more sensitive to correlated dephasing due to the larger number of qubits involved, highlighting the benefit of employing decoherence-free subspaces in algorithmic implementations. 
As measurements in chemistry generally refer to energy differences as opposed to absolute values, it is common to translate the potential energy curves to their nominal reference value at large separation $R$ as illustrated in 
Figure~\ref{fig:H2-fringe-scan-results}.b. This depiction more clearly reveals the respective upshift in energy with respect to the calculated binding energy (or well depth) and the simultaneously occurring shift in the position of the energy minimum towards larger internuclear distances. 

\begin{figure*}[t!] 
   \centering
   \includegraphics[scale=1]{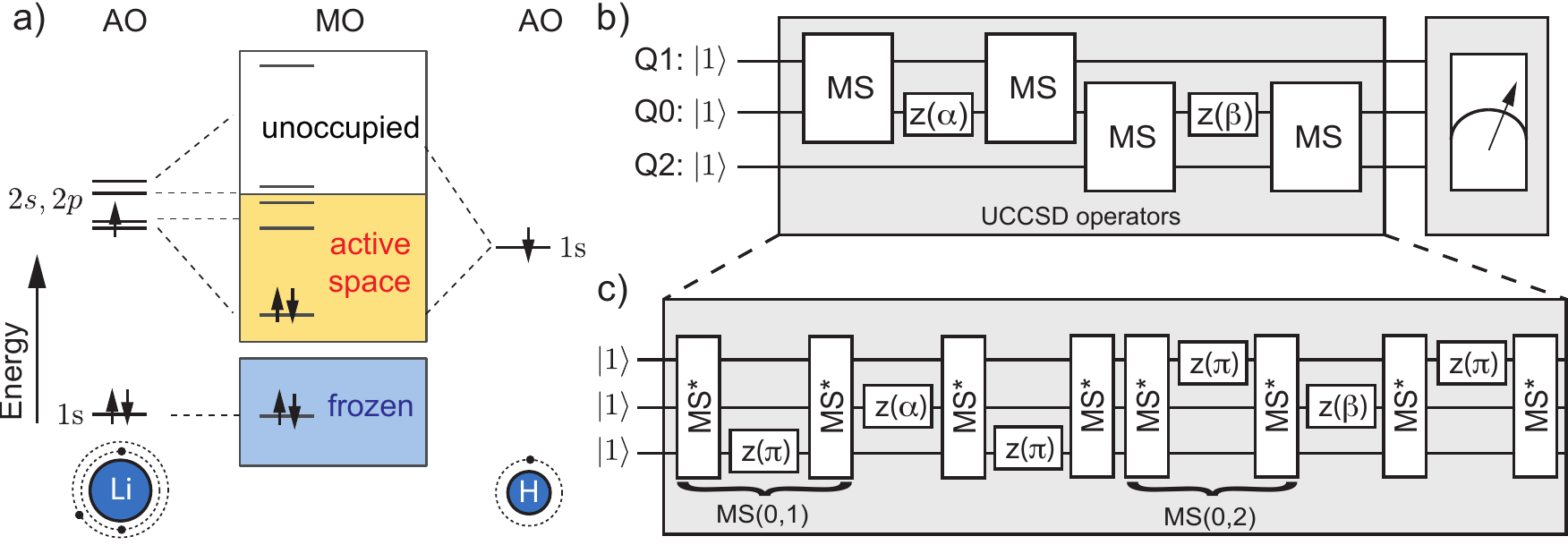} 
   \caption{\textbf{a)} LiH molecular orbitals (MO) formed out of atomic orbitals (AO) contributed by each element. The active space in which the implemented UCC excitation operators act is highlighted in yellow. \textbf{b)} Abstract quantum circuit implementing each of the two target UCCSD unitary operators on the the qubits indexed Q0, Q1 and Q2. A fully entangling gate (MS) acts locally between the enclosed qubits and surrounds a local qubit rotation z quantified by parameters $\alpha$ and $\beta$, respectively. \textbf{c)} Actual gate circuit implemented in the experiment. Each MS gate is broken up into global half-entangling gates (MS$^*$) that are interleaved with an addressed local $\pi$ phase shift, effectively restricting the entangling operation to the non-addressed qubit subset indicated below, e.g. MS(0,1) entangling only qubits 0 and 1.}
   \label{fig:LiHconcept}
\end{figure*}

We proceed to implement the VQE algorithm in full at five different internuclear separations $R$, yielding the results shown in Figure~\ref{fig:H2-fringe-scan-results}.c. For each configuration $R$, we start at a random initial value of $\theta_0$, prepare $\ket{\varphi_\mathrm{BK} (\theta_0)}$, measure the expectation values $\expval{H_\ell(\theta_0)}$ corresponding to the terms in the Hamiltonian and pass the measurement results to a Nelder-Mead simplex algorithm running on a classical computer. Here, the energy is calculated according to Eq.~\eqref{eq:energy} before a new value for $\theta$ is suggested for the next iteration. In parallel, we execute a noise-free simulation of the circuit at each iteration in order to monitor the convergence towards the theoretically expected energy. 

Generally, the algorithm converges in simulation and experiment. Residual energy fluctuations seen in the experimental results (cf. Figure~\ref{sfig:H2VQE}.a) are related to (1) measurement errors and (2) noisy gate operations. The first source of error stems from quantum projection noise (QPN)~\cite{Itano:1993} that scales proportional to $\sqrt{\nicefrac{1}{r}}$ for $r$ repetitions of the circuit (here, $r\leq1000$ for the VQE points). The second source of error is related to the experimental environment and manifests itself, e.g., in laser intensity fluctuations and transient electrical noise coupling to the motion of the ions. Both introduce a loss of fidelity in digital quantum simulations as shown in our earlier work \cite{Lanyon:2011} and can be mitigated through technical improvements.

As a result, we did not fix the number of VQE iterations to a specific value, but instead implemented a sinusoidal fit to the 1D parameter space of the energy explored throughout all iterations (cf.  Figure~\ref{sfig:H2VQE}.b), with each point weighted according to the QPN contributions from its constituent expectation value measurements. Figure~\ref{fig:H2-fringe-scan-results}.c shows each run's result superimposed on the previously discussed parameter scan. Error bars for the VQE points are derived from the above fitting procedure. In some cases, e.g. $R=0.6$ shown in the Figure~\ref{fig:H2-fringe-scan-results}.c inset, the simplex algorithm appears to get stuck, which likely is the result of a premature reduction in the step-width of $\theta$ caused by the noise sources discussed above. We return to this effect in the next section.

\section{Lithium Hydride}
We now increase the complexity by turning to a heteronuclear molecule with four electrons and aim to simulate the ground state energy of lithium hydride (LiH). This requires the introduction of additional variational parameters and thereby increases the circuit depth. LiH is also a natural small-molecule example and was previously simulated using four superconducting qubits in~\cite{Kandala:2017}. We implement its simulation using three ion qubits.

\subsection{Encoding the problem}

We again begin with the classical preprocessing step, choosing a minimal basis set of Slater-type orbitals represented by linear combinations of 6 Gaussian functions (STO-6G). A Hartree-Fock calculation then leads us to determine an energy-ordered molecular orbital basis into which the molecules' 4 electrons are filled sequentially (Figure~\ref{fig:LiHconcept}.a). The corresponding complete UCC ansatz, truncated to single and double excitations (UCCSD), yields 32 single and 168 double excitation operators. Under a Bravyi-Kitaev transformation, a direct implementation of this ansatz would require 12 qubits. Using the same approximation that was employed in the four qubit implementation of \cite{Kandala:2017}, we reduce the number of required qubits by identifying an active space of two electrons in three spatial orbitals, making an arbitrary choice in the degenerate subspace. Such active spaces average out (freeze) the core electrons not thought to be involved strongly in the correlations responsible for bonding \cite{Roos:1980}. 

Lastly, we take an additional step in order to identify the dominant contributions to the active space at an efficient level of classical theory known as configuration interaction singles and doubles (CISD). In this case, two singlet excitations are found to dominate and their corresponding unitary coupled cluster formulation is approximated as
\begin{equation*}
U_\mathrm{UCCSD}=e^{\alpha (a_5^{\dag}a_4^{\dag}a_3a_2-a_2^{\dag}a_3^{\dag}a_4a_5)} \cdot e^{\beta (a_7^{\dag}a_6^{\dag}a_3a_2-a_2^{\dag}a_3^{\dag}a_6a_7)},
\end{equation*}
where $\alpha, \beta$ are two components of the vector $\vec{\theta}$ and correspond to the parameters of the variational optimization. 

\begin{figure*}[t!]
   \centering
   \includegraphics[scale=1]{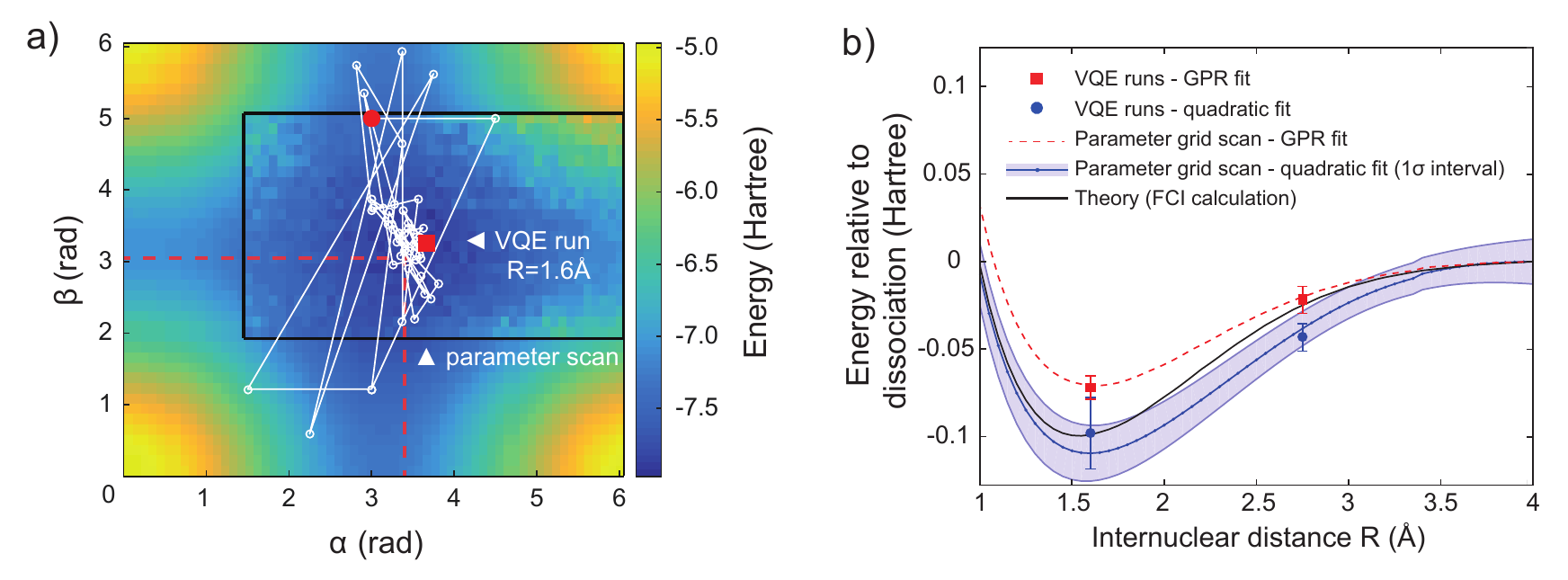} 
   \caption{LiH results.  
   \textbf{a)} Energy landscape at internuclear distance \mbox{$R=1.6~\mathrm{{\AA}}$}. The experimentally scanned parameter range, indicated by the black box inset, is superimposed on a theoretical calculation. The dashed red lines mark the coordinates of the targeted energy minimum. Connected white lines show the steps taken by the VQE algorithm, with the starting point marked by a filled red dot and the terminus by a red square.
   \textbf{b)} The theoretical LiH potential energy surface calculated for the minimal basis set (black) is shown in comparison with experimentally obtained results, offset to overlap at maximum distance $R$ to better illustrate the well-depth differences in the grid scan reconstructions. In absolute values, all measured data is above the FCI calculation. The data points result from sampling the energy landscape $\expval{H(R)}_{\alpha, \beta}$ using the VQE algorithm or the parameter scan shown in a) and fitting the explored space with a Gaussian process regression (GPR) based machine learning algorithm (red dashed line) or a 2D quadratic fit (blue solid line). The error bars are obtained from the fits with the underlying data weighted by quantum projection noise. The slight \enquote{kink} close to $R=3.5~\mathrm{{\AA}}$ is due to the interplay of rounding errors introduced in the fitting routine with small deviations originating in our active space approximation.}
   \label{fig:LiH}
\end{figure*}

After mapping the above fermionic representation to Pauli operators using the Bravyi-Kitaev (BK) transformation, only three qubits are acted on non-trivially, i.e. in a way that changes state populations. This is related to the previously observed fact that BK computational basis states naturally reflect certain spin symmetries, allowing for a more compact representation in some cases than the corresponding Jordan-Wigner mapping \cite{OMalley:2016,Bravyi:2017}. The problem may hence be reduced to its action on these three qubits and simulated by selecting appropriate subterms, acting on the initial state \mbox{$\ket{\varphi(0)} = \ket{111}$}. 
The final operator we implement has the form
\begin{equation}
U^\mathrm{BK}_\mathrm{UCCSD} (\alpha, \beta) = e^{-i\alpha \sigma_2^x\sigma_4^y} \cdot  e^{-i\beta \sigma_2^x\sigma_6^y}. 
\label{eq:BKUCC}
\end{equation}
Further details of this derivation can be found in Appendix~\ref{sup:LiH}.
The quantum circuit corresponding to Eq.~\eqref{eq:BKUCC}, which acts on subsets of the qubit register, is shown in Fig.~\ref{fig:LiHconcept}.b. We choose to implement it in the experiment using the circuit shown in Fig.~\ref{fig:LiHconcept}.c, which is based on a refocusing technique \cite{Muller:2011}. Here, an addressed $\pi$-phase shift between two half-entangling MS gates effectively decouples the addressed qubit from the two remaining qubits: these become entangled, while the phase-shifted qubit does not obtain any correlations with the rest of the register. Other implementation strategies range from other algorithmic solutions such as spectroscopic decoupling~\cite{Schindler:2013} or sequence recompilation~\cite{Martinez:2016} to hardware solutions based on Raman gates \cite{Debnath:2016}.

\subsection{Results}
We first perform a parameter scan to establish a baseline for the performance of our system before implementing the VQE algorithm for select points. Moving sequentially over a section of the two-dimensional parameter space bound by $\alpha = [1.5, 6 ]$, $\beta = [2, 5]$ in a grid-like pattern, we perform three rounds of projective measurements at each setting to determine expectation values for each term in the BK transformed Hamiltonian: $\expval{H_\ell} = \{{Z_0}$, ${Z_1}$, ${Z_2}$, ${Z_1Z_0}$, ${Z_2Z_0}$, ${Z_2Z_1}$, ${X_1X_0}$, ${Y_1Y_0}$, ${X_2X_0}$, ${Y_2Y_0}$, ${X_2X_1}$, ${Y_2Y_1}\}$ (see Fig.~\ref{fig:LiHscan_expval} for data).
The results are combined via Eq.~\eqref{eq:energy} in order to calculate the energy landscape for each internuclear separation $R$. An example for $R = 1.6\,$\AA~is shown in Figure~\ref{fig:LiH}.a) with the experimentally measured parameter space superimposed on a theoretical calculation of the full range. 

In order to reconstruct the full potential energy curve of the electronic ground state from this data, we investigate two approaches: (1) a two-dimensional quadratic fit to the energy minimum and (2) a Gaussian process regression (GPR) fit. The fit minima from all energy surfaces of the different internuclear separations $R$ finally yield the potential energy curves in Fig.~\ref{fig:LiH}.b, which we again use as experiment-based reference.

We now implement an iterative VQE procedure similar to the one described in the case of H$_2$ above, but with an important modification. Numerical simulations, detailed in Appendix~\ref{sup:VQELiH}, and experiments reveal that the previously observed convergence failure of the bare Nelder-Mead search algorithm in the presence of noise has a significant impact on the accuracy of the energies obtained in each VQE run. To combat this effect, we switch the optimization to a hybrid algorithm~\cite{Vugrin:2005} that also incorporates an element of simulated annealing by introducing random perturbations, sampled from a distribution $D$, that are added to the cost function in Eq.~\eqref{eq:energy}. 
In this way, the VQE algorithm is forced to continuously sample the surroundings of the minimum as shown in Fig.~\ref{fig:LiH}.a without converging any further. The larger number of samples effectively allows for a more precise estimation of the minimum's location through a fit to the data. 
We heuristically choose $D$ to be on the order of the energy error caused by quantum projection noise such that the perturbations only become dominant in the vicinity of the minimum. For example, at \mbox{$R = 1.6~\mathrm{\AA}$} the energy error from quantum projection noise after 500 repetitions of the experiment is between \mbox{$0.01~\text{Ha} \le \Delta\expval{H} \le 0.04~\text{Ha}$}, depending on the specific parameter set $\{\alpha, \beta\}$. Hence we chose to sample from the uniform distribution \mbox{$D = [0.01 , 0.08]~\text{Ha}$}, with mean \mbox{$\bar{D}=0.045~\text{Ha}$}, comparable to the range above. Once the fluctuations of the energy values in the VQE execution are on the order of the mean perturbation strength $\bar{D}$, we proceed for another $10-20$ iterations before stopping the outer VQE loop. We then evaluate the data using (1) a two-dimensional quadratic fit to a subset of the VQE iterations (four standard deviations from the median; illustrated in Fig.~\ref{fig:LiH_3D}) or (2) a Gaussian process regression fit. Both sets of results are shown in Fig.~\ref{fig:LiH}.b in comparison with the ideal theoretical result for our basis set.
While resulting in a smooth potential energy surface, the GPR-based fit appears to systematically underestimate the binding energy, which highlights the impact the chosen data evaluation method has in the final step of a VQE algorithm.

\begin{figure}[!t] 
   \centering
   \includegraphics[scale=0.6]{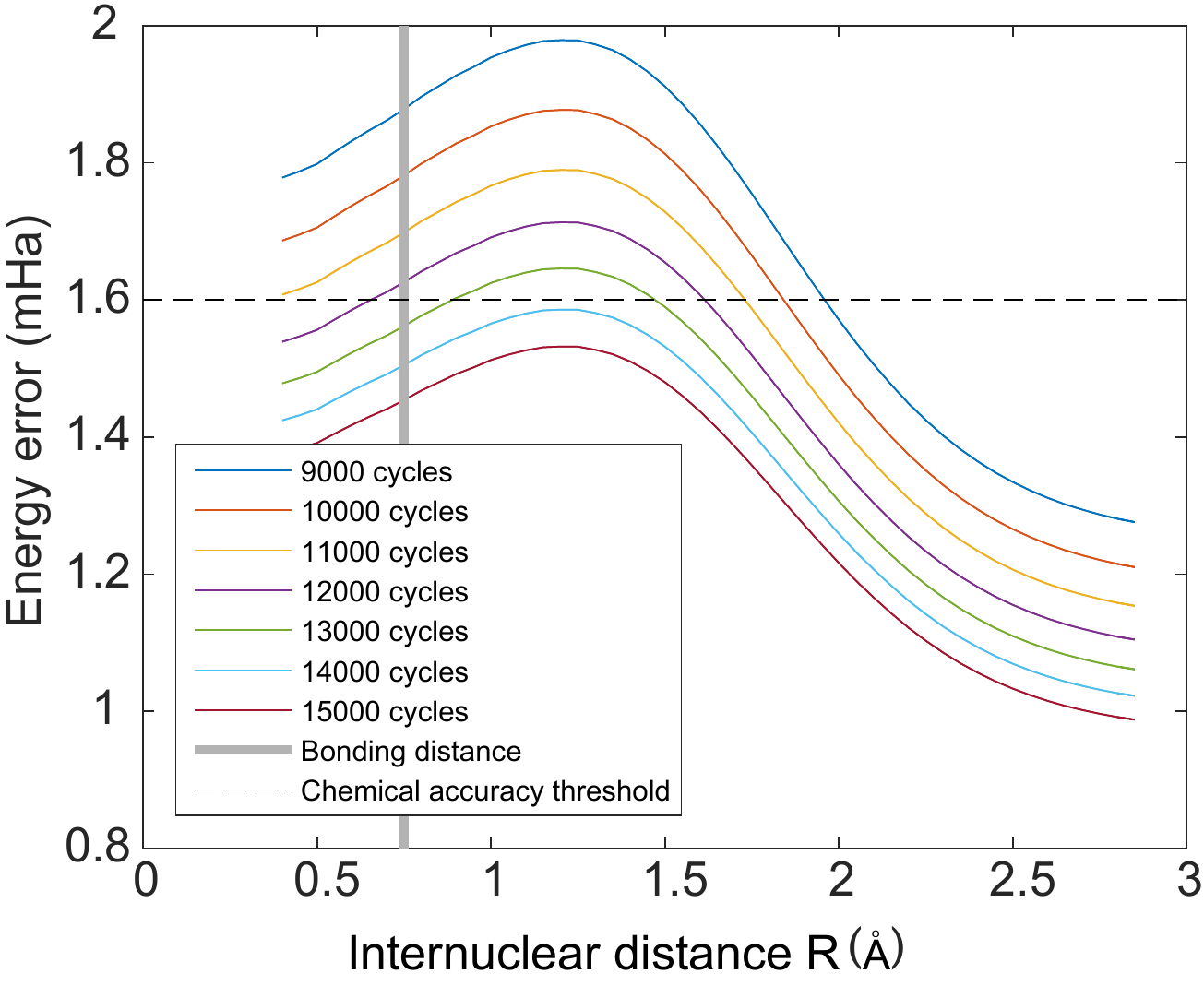} 
   \caption{Scaling of the quantum projection noise related energy error in H$_2$. Each color-coded curve corresponds to a $1\sigma$ error in the energy calculated from the Hamiltonian's expectation values $\expval{H_\ell}$ appropriately weighted by the scalars $c_\ell(R)$ associated with the respective internuclear distances $R$.}
   \label{fig:errorVScycles}
\end{figure}

\section{Discussion}
To put the results presented above into perspective, it is useful to establish a benchmark against which they can be compared. For quantum chemistry calculations, the widely used concept of \enquote{chemical accuracy} constitutes such a point of reference.

Chemists are often particularly interested in free energy landscapes which provide mechanistic insight into chemical events of significant practical importance such as drug binding, catalysis and material properties.
Free energies are obtained from partition functions which can be sampled using Monte Carlo or Molecular Dynamics simulations, which assume the ability to compute solutions to the electronic structure problem. The resulting energy landscapes must be extremely accurate as chemical rates are exponentially sensitive to changes in free energy and thus changes in potential energies. This sensitivity can be seen from the Erying equation~\cite{Eyring:1935} for chemical rates which is proportional to 
$\nicefrac{e^{-\beta\Delta G^\ddag}}{\beta},$
where $\Delta G^\ddag$ is the difference in free energy between reactants and the transition state, and $\beta$ is the inverse temperature in atomic units. At room temperature and atmospheric pressure, an energy error in $\Delta G^\ddag$ of \mbox{$43.3 \times 10^{-3}$~eV} \mbox{($1.6\times 10^{-3}$~ Hartree)} translates to a chemical rate that is wrong by a factor of ten \cite{Helgaker:2002}. 

With respect to this threshold, it follows that the expectation values of most operators $\expval{H_\ell}$ in the chemical Hamiltonian need to be determined at a similar level of precision. In practical terms, this translates to a minimum number of measurement averages required to overcome the intrinsic quantum projection noise in the estimation of each Pauli string $H_\ell$.  Figure~\ref{fig:errorVScycles} illustrates this requirement for the case of the $\mathrm{H}_2$ molecule, assuming no other sources of noise and knowledge of the final value of each constituent expectation value at every point.

The graph shows that one would need to repeat each of the measurements at least 15000 times in order to reduce the intrinsic measurement noise below chemical accuracy. Given that every repetition incurs the overhead of state initialization, state preparation and measurement, the number of repetitions has a strong impact on run-time. For example, on our trapped ion system the time needed for one repetition is on the order of 20~ms, requiring at least 5 minutes of averaging to ensure a measurement limit at chemical accuracy in each VQE iteration. In the current setup this limitation stems from both state initialization through laser cooling and the duration of quantum gate operations. However, ways to overcome these constraints have been demonstrated in a variety of experiments already \cite{Lechner:2016,Schafer:2018, WongCampos:2017}, promising order-of-magnitude advances in speed and thereby much shorter runtimes. 
In addition to technical improvements on the hardware side, an adaptive measurement strategy might also alleviate the resource needs with respect to the required number of averages. This could entail varying the number of repetitions throughout the VQE iterations based on the observed energy changes in each step, gradually increasing the measurement precision as the algorithm converges towards the energy minimum. 

\begin{figure}[!t] 
   \centering
   \includegraphics[scale=1]{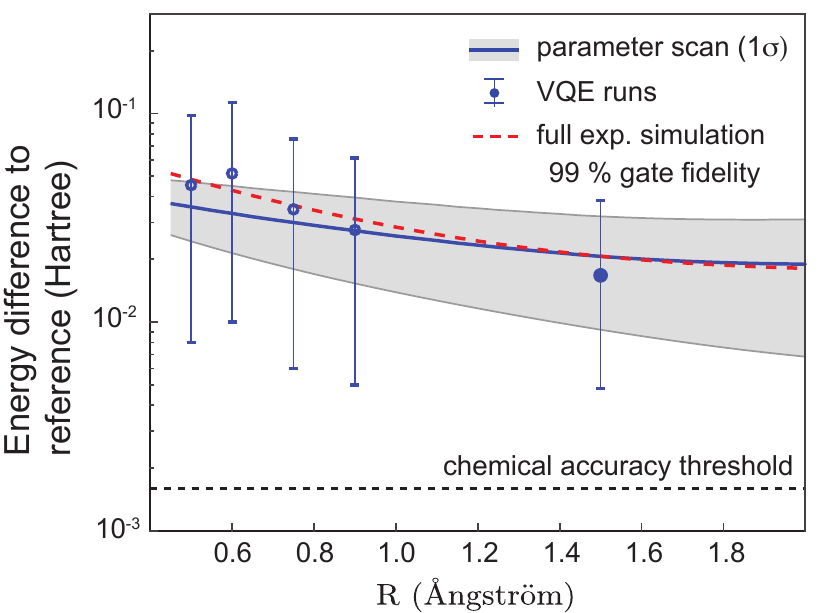}
   \caption{Energy errors of the reconstructed H$_2$ potential energy surface and the influence of decoherence. Differences are given with respect to the full configuration interaction (FCI) calculation performed in the chosen molecular basis. The red line corresponds to a full simulation of the quantum circuit, including multiple decoherence channels and the experimentally determined gate fidelity (see Appendix~\ref{sub:H2decoherenceSim} for details).}
   \label{fig:energy_errors}
\end{figure}

In the proof-of-principle experiments above we have focused on probing the effects of system noise from both a limited number of measurements and decoherence effects rather than achieving chemical accuracy. The parameter scans for H$_2$ (LiH) were taken by averaging 100 (500) repetitions of the corresponding circuit, which for the case of LiH already corresponded to a significant time overhead due to the two-dimensional nature of the parameter space. As a consequence, there is a large uncertainty associated with the energy error of the extracted potential energy curves. 
The effect of errors resulting from decoherence, seen in Figure~\ref{fig:H2-fringe-scan-results}.a, was modeled for the case of molecular hydrogen using the Bravyi-Kitaev transformation with help of the quantum chemistry package OpenFermion \cite{McClean:2017b}. The result is shown in Figure~\ref{fig:energy_errors} and largely matches the upshift in energy observed in the experiment.

The VQE method appears to be robust to calibration errors in some of the gates as evidenced by the shift in the parameter value $\theta$ in Fig.~\ref{fig:H2-orb-circ-expval}.c, also seen in \cite{OMalley:2016}. However, similar to standard quantum tomography, the method still relies on accurate determination of the measurement operators $\expval{H_\ell}$, which in turn are bound to local gate operations that rotate the desired basis for each qubit into the measurement basis. Errors at this point in the circuit will be folded into the results and cannot directly be recovered from. Possible ways around this issue might be found both on the quantum and classical side. Quantum control techniques tailored to experimental errors allow for the suppression of certain types of gate errors~\cite{Merrill:2014}. An algorithmic modification might be the addition of a second variational optimization loop parametrizing the pre-measurement rotations, which is entered once the nominal VQE outer loop has converged to a desired level. In this way, the measurement basis can be corrected for offsets in the same way as the calibration errors inside the quantum circuit. Lastly, the quantum circuit might be extended to include ancilla qubits, which allow for error detection based on stabilizers that indicate whether the computational subspace was still maintained at circuit completion~\cite{Linke:2017}. In this way, post selection might be used to raise the effective fidelity of non-error corrected quantum devices to a level that makes them useful for quantum chemistry calculations in the near-term future.

\section{Conclusion}
In summary, we have performed the first multi-ion quantum simulation of quantum chemistry. In our simulation of molecular hydrogen, we have employed both the Bravyi-Kitaev as well as the Jordan-Wigner transformation in the mapping from fermions to qubits and varied the respective Hartree-Fock reference state to reveal the impact of decoherence. Unstable behavior of a classical Nelder-Mead optimization routine was circumvented in our simulation of lithium hydride by amending the variational optimization with simulated annealing and subsequently employing a quadratic fit to estimate the location of the minimum in the two-dimensional energy landscape. When scaling up to larger parameter spaces, this step of the classical data evaluation routine will likely become an even more important issue.

Despite a significant increase in experiment runtime, it was still possible to sequentially scan the two-dimensional parameter space of LiH. However, the addition of additional excitation operators for more accurate energy determinations and the scaling up to more complex molecules reveals the true power of the VQE approach: sparse and adaptive sampling of a potentially very high dimensional energy landscape. While the number of different measurements needed to estimate the energy of an arbitrary molecule can be large, recent work shows how using structure fundamental to the fermionic nature of particles can reduce this number by an order of magnitude or more for minimal additional complexity \cite{Rubin:2018}.

To deliver on the promise of near-term usefulness, further work is needed on both the quantum and classical aspects of the VQE algorithm. The mitigation of errors in short depth quantum circuits has increasingly come into focus \cite{Temme:2016,McClean:2017,Colless:2017,Johnson:2017,Endo:2017,Rubin:2018} and the use of dynamic error suppression techniques at the physical gate level promises improved fidelities of the practically implemented operations. A recent numerical study~\cite{Otten:2018} specifically investigates the impact of errors in the simulation of the ground state energy of H$_2$ and LiH and illustrates the expected impact on reaching chemical accuracy.

Furthermore, there are a number of recent proposals in the literature which appear to significantly reduce the resources required to scale up these simulations. For instance, in \cite{Babbush:2017} authors introduce a new class of basis functions for the simulation of electronic structure. They quadratically reduce the number of terms in the Hamiltonian from ${\cal O}(N^4)$ to ${\cal O}(N^2)$, which also reduces naive bounds on the number of measurements from ${\cal O}(N^8)$ to ${\cal O}(N^4)$ \cite{Wecker:2015}. This speeds up data collection by making the number of circuit repetitions substantially more practical. The number of measurements required can also be improved further using recent techniques for enforcing $n$-representability conditions (known constraints on the geometry of fermionic states) from \cite{Rubin:2018}. 

The most significant challenge in realizing variational algorithms in the near-term remains the circuit depth required. A naive application of unitary coupled cluster requires a number of gates that scales as ${\cal O}(N^4)$ assuming arbitrary connectivity between qubits. This strongly suggests that more scalable variational circuits will be needed if we are to approach classically intractable calculations without error-correction. One alternative to unitary coupled cluster is to use an unstructured variational circuit, as demonstrated recently in \cite{Kandala:2017}. Recent work in \cite{McClean:2018}, however, has shown that such strategies become exponentially expensive as a consequence of concentration of measure in random quantum circuits. Nevertheless, there has been significant recent progress in realizing low depth structured variational ansatze; e.g., in \cite{Kivlichan:2018} the authors introduce variational circuits based on Trotterized adiabatic state preparation which can be implemented with linear gate depth even on a device with extremely limited (e.g. next-neighbor) qubit connectivity. In \cite{DallaireDemers:2018}, a similar ansatz is introduced, also with linear gate depth which is equivalent to the fermionic swap network from \cite{Kivlichan:2018} with a different variational parameter initialization.

It is far from a forgone conclusion that we will one day solve classically intractable problems in quantum chemistry without error-correction. Yet, by using these ever improving algorithms combined with error-mitigation strategies and simpler Hamiltonian representations, it is certainly the case that we should be able to push forward and obtain reasonably accurate quantum simulations of molecules with tens of spin-orbitals (qubits) in the near future.
In comparison to other quantum simulations with less stringent requirements, the notion of chemical accuracy in this context provides a clear benchmark to determine the point at which support by quantum processors transitions into a useful regime. In this way, metrics like the deviation in well-depth or the worst case deviation along the entire potential energy surface, captured by the so-called non-parallel error, might in fact provide a convenient way to measure multi-qubit performance in different architectures, which will help assess technological progress.

\section*{Acknowledgments}

This work was supported by the Austrian Science Fund (FWF) under Grants No. P25354-N20 and F4002-N16 (SFB FoQus), by the European Commission via the integrated project SIQS and the Institute f\"ur Quanteninformation GmbH. The research is further based upon work supported by the Office of the Director of National Intelligence (ODNI), Intelligence Advanced Research Projects Activity (IARPA), via the U.S. Army Research Office Grant No. W911NF-16-1-0070. The views and conclusions contained herein are those of the authors and should not be interpreted as necessarily representing the official policies or endorsements, either expressed or implied, of the ODNI, IARPA, or the U.S. Government. The U.S. Government is authorized to reproduce and distribute reprints for Governmental purposes notwithstanding any copyright annotation thereon. Any opinions, findings, and conclusions or recommendations expressed in this material are those of the author(s) and do not necessarily reflect the view of the U.S. Army Research Office. CH further acknowledges support by the ARC Centre of Excellence for Engineered Quantum Systems (CE110001013). PJL and JR were supported by AFOSR award no. FA9550-12-1-0046. AAG acknowledges support from the Army Research Office under Award No. W911NF-15-1-0256 and the Vannevar Bush Faculty Fellowship program sponsored by the Basic Research Office of the Assistant Secretary of Defense for Research and Engineering (Award number ONR 00014-16-1-2008).

\appendix

\section*{Appendix}
\section{Qubit implementation}
\label{ap:qubit_implementation}
\begin{figure}[!t]
\begin{center}
\includegraphics[scale=1]{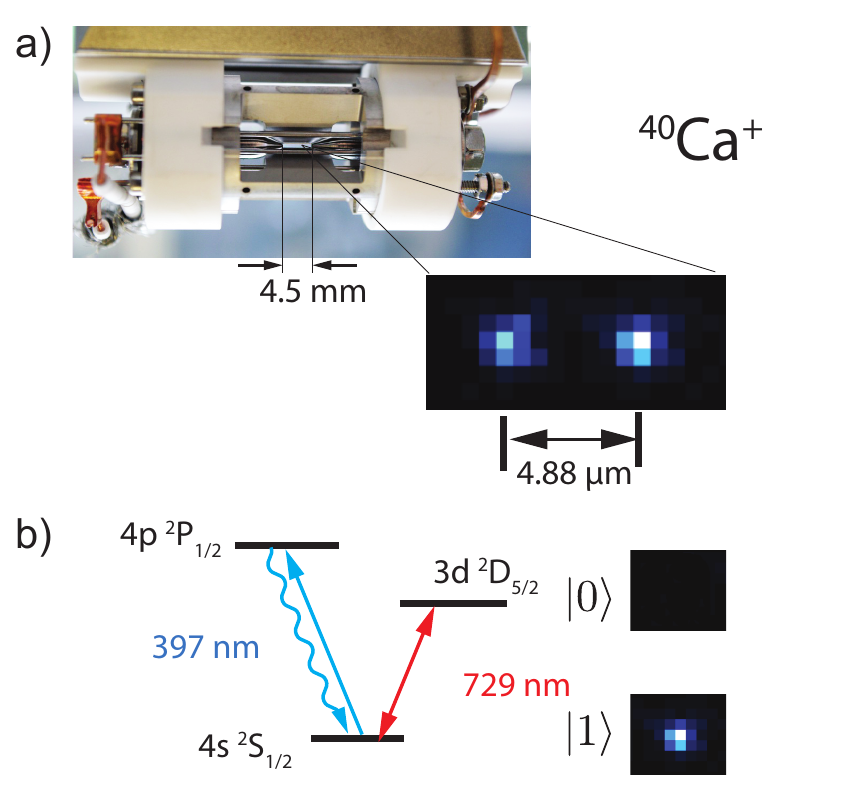}
\caption{Qubit implementation using trapped ions \textbf{a)} Linear Paul trap with ion-electrode distance of 565~$\upmu$m capable of stably storing linear arrays up to 70 ion qubits (up to 20 fully controlled). The trap is housed in an ultra-high vacuum environment, connected to DC and AC voltage sources that provide the confining electric fields and is specifically designed to provide laser beam access for the manipulation of the individual ions. \textbf{b)} Simplified level scheme of $^{40}$Ca$^+$. A single-photon transition at 729~nm is used to manipulate the qubit encoded in two Zeeman states of the S and D manifolds. Upon illumination from a laser at 397~nm, the qubits are projected into either the \ket{1} state, yielding ion fluorescence, or the \ket{0} state with no observable ion fluorescence.}
\label{fig:qubit_implementation}
\end{center}
\end{figure}

The qubits used for this experiment are encoded in Zeeman sublevels of a metastable D and an S ground state of $^{40}$Ca$^+$ ions confined in a macroscopic linear Paul trap described in detail in \cite{Hempel:2014} and shown in Figure~\ref{fig:qubit_implementation}.a. 
In particular, we associate 
\begin{align*}
\ket{0} =& \ket{\uparrow} = \ket{\mathrm{3d\,^2D}_{5/2}\,(m=+3/2)}\quad \text{and}\\
\ket{1} =& \ket{\downarrow} = \ket{\mathrm{4s\,^2S}_{1/2}\,(m=+1/2)}. 
\end{align*}
Both electronic levels are connected via an optical quadrupole transition at 729~nm, which is used to drive the qubits via an ultra-stable laser with $\sim1$~Hz linewidth. The same laser mediates multi-qubit interactions via the joined motional modes of the ions in their trapping potential. An auxiliary transition at 397~nm between $\mathrm{4s\,^2S}_{1/2}$ and $\mathrm{4p\,^2P}_{1/2}$ is used for laser cooling, state preparation via optical pumping and state detection. 

Each experimental repetition consists of laser cooling, ground state initialization (sideband cooling to the motional ground state of the ion strings center of mass mode and optical pumping to the electronic state \ket{1}), execution of the desired gate sequence and finally parallel state detection for each qubit via an electron multiplying CCD camera (Figure~\ref{fig:qubit_implementation}.b). Every sequence is repeated at least 100 times to calculate an expectation value from the observed probabilities. 

\section{Molecular Hamiltonian for H$_2$ and UCCSD operators}
\label{ap:UCCH2}

\subsection{Bravyi-Kitaev (BK) transformation}
The molecular hamiltonian for H$_2$ in the STO-3G minimal basis can be mapped from its fermionic form \eqref{eq:ferm_hamiltonian} to qubits yielding
\begin{align}
H^\mathrm{BK} = & f_0\mathcal{I} + f_1 \sigma^{z}_{0} + f_2 \sigma^{z}_{1} + f_3 \sigma^{z}_{2} + f_4 \sigma^{z}_{1} \sigma^{z}_{0} + f_5 \sigma^{z}_{2} \sigma^{z}_{0} \notag \\ 
&+ f_6 \sigma^{z}_{3} \sigma^{z}_{1} + f_7 \sigma^{x}_{2} \sigma^{z}_{1} \sigma^{x}_{0} + f_8 \sigma^{y}_{2} \sigma^{z}_{1} \sigma^{y}_{0} + f_9 \sigma^{z}_{2} \sigma^{z}_{1} \sigma^{z}_{0} \notag \\ 
 &+ f_{10} \sigma^{z}_{3} \sigma^{z}_{2} \sigma^{z}_{0} + f_{11} \sigma^{z}_{3} \sigma^{z}_{2} \sigma^{z}_{1} + f_{12} \sigma^{z}_{3} \sigma^{x}_{2} \sigma^{z}_{1} \sigma^{x}_{0} \notag \\
 &+ f_{13} \sigma^{z}_{3} \sigma^{y}_{2} \sigma^{z}_{1} \sigma^{y}_{0} + f_{14} \sigma^{z}_{3} \sigma^{z}_{2} \sigma^{z}_{1} \sigma^{z}_{0}
\label{hamiltonianBK}
\end{align}
where the coefficients $f_i$ depend on the internuclear separation ($R$) between the hydrogen atoms and are derived from the integrals in \eqref{eq:integrals}.

The reference state for the calculation corresponds to the Hartree-Fock solution  \mbox{$|\varphi_\mathrm{HF}\rangle= |0001 \rangle$}. We observe that the terms in the Hamiltonian only act with the identity and $\sigma^{z}$ operations on qubits 1 and 3. This fact allows us to rewrite our reference state as \mbox{$|\varphi_\mathrm{HF}\rangle= |0 \rangle _1 |0\rangle_3 \otimes |0\rangle_2 |1\rangle_0$}. As qubits 1 and 3 will not experience population changes under the Hamiltonian, we can reduce Eq.~\eqref{hamiltonianBK} to an effective Hamiltonian acting on two qubits, with reference state $|\varphi_{HF}\rangle= |01 \rangle$:
\begin{align}
H^\mathrm{BK}=&c_0 \mathcal{I} + c_1\sigma^{z}_{0}+c_2 \sigma^{z}_{1}\notag \\ &+ c_3 \sigma^{z}_{0}\sigma^{z}_{1}+ c_4 \sigma^{x}_{0}\sigma^{x}_{1} + c_5 \sigma^{y}_{0}\sigma^{y}_{1}.\nonumber
\end{align}
Here we have relabeled qubits 0 and 2 as 0 and 1. The coefficients $c_i$ are now given by:
\begin{align}
c_0&=f_0+f_2+f_6& c_3&=f_5+f_9+f_{10}+f_{14}\notag\\ 
c_1&=f_1+f_4&  c_4&=f_7+f_{12}\notag\\ c_2&=f_3+f_{11}& c_5&=f_8+f_{13} \notag  
\label{coefficients}
\end{align}
This reduction of the problem for the hydrogen molecule was first noted in~\cite{OMalley:2016}, developed into a general method in~\cite{Bravyi:2017} and used in superconducting implementations of several problems in~\cite{Kandala:2017}.

\subsection{Jordan-Wigner (JW) transformation}
The molecular Hamiltonian under the Jordan-Wigner transformation gets mapped to 

\begin{align}
H^\mathrm{JW}=&c_0\mathcal{I}+c_1(\sigma^{z}_{0}+\sigma^{z}_{1})+c_2(\sigma^{z}_{2}+\sigma^{z}_{3}) + c_3\sigma^{z}_{3} \sigma^{z}_{2} \notag \\
&c_4 \sigma^{z}_{1} \sigma^{z}_{0}+ c_5(\sigma^{z}_{2} \sigma^{z}_{0} + \sigma^{z}_{3} \sigma^{z}_{1})+ c_6(\sigma^{z}_{2} \sigma^{z}_{1}+ \sigma^{z}_{3} \sigma^{z}_{0}) \notag \\
&+c_7(\sigma^{x}_{3} \sigma^{y}_{2}\sigma^{y}_{1} \sigma^{z}_{0}+\sigma^{y}_{3} \sigma^{x}_{2}\sigma^{x}_{1} \sigma^{y}_{0})\notag\\
&-c_7(\sigma^{x}_{3} \sigma^{x}_{2}\sigma^{y}_{1} \sigma^{y}_{0}+\sigma^{y}_{3} \sigma^{y}_{2}\sigma^{x}_{1} \sigma^{x}_{0}),
\end{align}
with coefficients $c_i$ again derived from the integrals \eqref{eq:integrals}. 

Under the Jordan-Wigner transformation all the qubits are used to store occupation numbers while in in the Bravyi-Kitaev transformation even qubits store occupations and odd qubits keep track of the parity of all the qubits with smaller indices. Hence, four qubits are needed to encode the ansatz state \mbox{$|\varphi_\mathrm{HF}\rangle= |0011 \rangle$}.

\subsection{Application of Unitary Coupled Cluster to H$_2$}

For H$_2$ in the minimal basis, the second quantized formulation of the Unitary Coupled Cluster operator for single and double excitations corresponds to
\begin{equation*}
U=\exp[\theta_{01}^{23}(a_2^{\dag}a_3^{\dag}a_1a_0-a_0^{\dag}a_1^{\dag}a_3a_2)],
\end{equation*}
where $\theta_{01}^{23}$ is the coupled cluster amplitude that is variationally optimized. Note that in this case the single-excitation operators are effectively incorporated in the basis we are using (i.e. the single excitations rotate the basis and do not need to be applied explicitly in the circuit). Using the BK mapping this operator is expressed as follows:
\begin{align}
&U(\theta_{01}^{23})= \exp \Bigg( i\frac{\theta_{01}^{23}}{8} \Big[-\sigma^{x}_{2}\sigma^{y}_{0}+\sigma^{y}_{2}\sigma^{x}_{0}-\sigma^{x}_{2}\sigma^{z}_{1}\sigma^{y}_{0}  
+\sigma^{y}_{2}\sigma^{z}_{1}\sigma^{x}_{0}\nonumber\\&-\sigma^{z}_{3}\sigma^{x}_{2}\sigma^{y}_{0} +\sigma^{z}_{3}\sigma^{y}_{2}\sigma^{x}_{0}-\sigma^{z}_{3}\sigma^{x}_{2}\sigma^{z}_{1}\sigma^{y}_{0}+\sigma^{z}_{3}\sigma^{y}_{2}\sigma^{z}_{1}\sigma^{x}_{0}\Big] \Bigg).\nonumber
\end{align}

As the terms in $U(\theta_{01}^{23})$ only act with the identity and $\sigma^{z}$ on qubits 1 and 3, the operator can be reduced to
\begin{align}
U(\theta_{01}^{23})&=\exp \left( i\frac{\theta_{01}^{23}}{2} [-\sigma^{x}_{1}\sigma^{y}_{0}+\sigma^{y}_{1}\sigma^{x}_{0}] \right)\notag \\&= \exp \left( -i\frac{\theta_{01}^{23}}{2}\sigma^{x}_{1}\sigma^{y}_{0} \right) \exp \left(i\frac{\theta_{01}^{23}}{2} \sigma^{y}_{1}\sigma^{x}_{0}\ \right), \notag
\end{align}
where we have relabeled qubit 2 as 1 and used the fact that the two operators commute. Finally, we observe that the two exponentials in $U(\theta_{01}^{23})$ perform the same operation when applied to the reference state $|01\rangle$, explicitly
\begin{align}
\exp &\left(i\frac{\theta_{01}^{23}}{2} \sigma^{y}_{1}\sigma^{x}_{0}\ \right) |01\rangle \notag\\& = \left(\cos(\frac{\theta_{01}^{23}}{2}) I + i \sin(\frac{\theta_{01}^{23}}{2})\sigma^{y}_{1}\sigma^{x}_{0} \right) |01\rangle \notag \\
& = \left(\cos(\frac{\theta_{01}^{23}}{2}) I + i \sin(\frac{\theta_{01}^{23}}{2})\sigma^{x}_{1}\sigma^{y}_{0} \sigma^{z}_{1}\sigma^{z}_{0} \right) |01\rangle \notag \\
& = \left(\cos(\frac{\theta_{01}^{23}}{2}) I - i \sin(\frac{\theta_{01}^{23}}{2})\sigma^{x}_{1}\sigma^{y}_{0} \right) |01\rangle \notag \\
& = \exp \left(-i\frac{\theta_{01}^{23}}{2} \sigma^{x}_{1}\sigma^{y}_{0}\ \right) |01\rangle \notag.
\end{align}

\noindent This allows us to define the ansatz via the $U(\theta_{01}^{23})$ operator simply as
\begin{align}
U(\theta_{01}^{23})=& \exp \left( -i \theta_{01}^{23} \sigma^{x}_{1}\sigma^{y}_{0} \right).
\label{UCC3}
\end{align}
Note that this form is only valid when the operator acts on the reference state $|01\rangle$. 

\subsection{Implementation of unitary coupled cluster operator using MS gates}

In order to implement the UCC operator \eqref{UCC3} above using M\o{}lmer-S\o{}rensen gates, we employ a technique first demonstrated in M\"{u}ller et al. \cite{Muller:2011}, formulae 10-12. If we consider arbitary tensor products of qubit Pauli operators $A$ and $B$ with $[A,B]\neq 0$ we have: 
\begin{equation*}
\exp{(-i\alpha A)} \exp{(i\theta B)} \exp{(i\alpha A)} = \exp {(i\theta B^\prime)}, 
\end{equation*}
with $B^\prime = \exp{(-i\alpha A)} B \exp{(i\alpha A)}$ and using the fact that Pauli operators are self-inverse:
\begin{equation*}
B^\prime = (\mathcal{I}\cos\alpha - i A\sin\alpha)B(\mathcal{I}\cos\alpha + i A\sin\alpha)
\end{equation*}
using the fact that $A$ and $B$ do not commute and therefore must anticommute we obtain:
\begin{equation}
B^\prime = B\cos 2\alpha -  \frac{i}{2}[A,B]\sin 2\alpha 
\nonumber
\end{equation}

\begin{figure}[!t] 
   \includegraphics[scale=1]{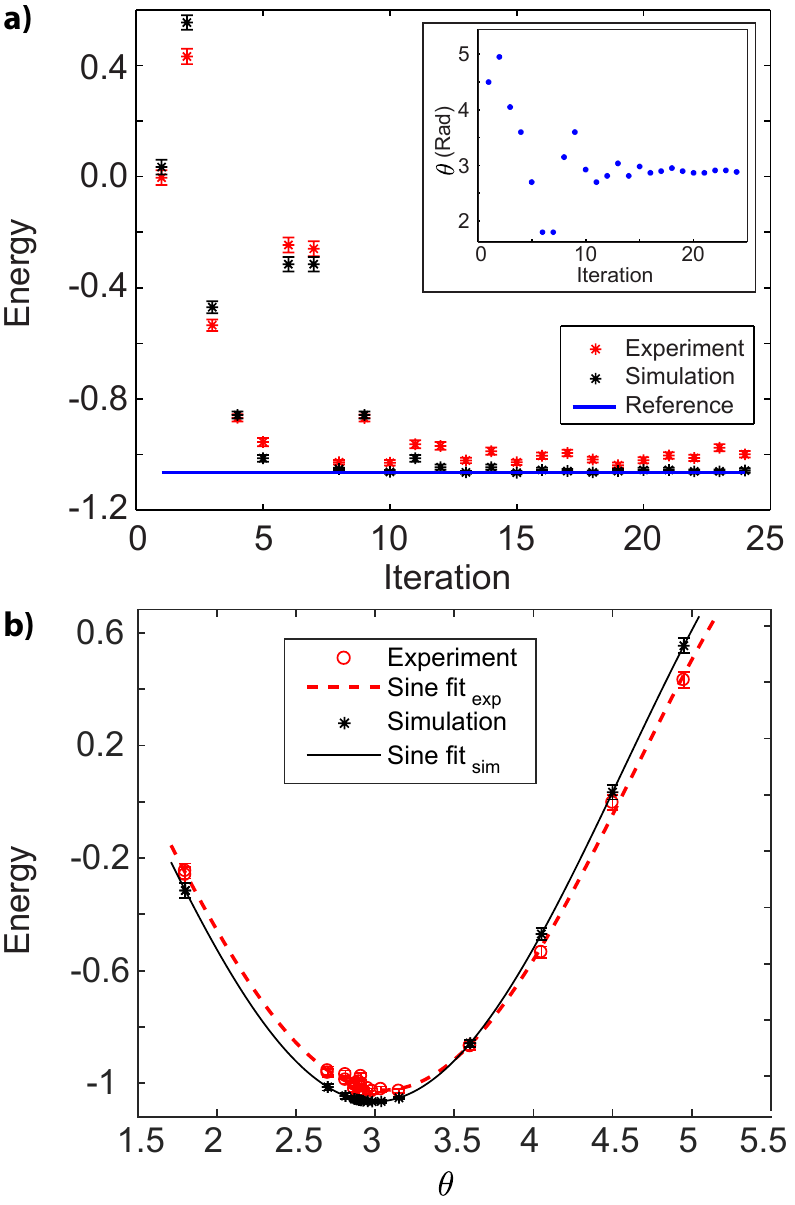} 
   \caption{Online VQE runs for $R=0.5$~\AA. \textbf{a)} Energy vs. Iteration number (target = blue line, simulated result = black, experimental result = red) Inset: Rotation angle $\alpha$ vs. Iteration number. Error bars derived from quantum projection noise. \textbf{b)} Energy vs. Rotation angle $\alpha$ with sinusoidal fitting (simulation = black, experiment = red).}
   \label{sfig:H2VQE}
\end{figure}

Specifically, for $\alpha = \pi/4$ and the case above 
\begin{align}
\exp(&-i\theta\sigma_y^1\sigma_x^j) = \sin{\left(i\theta\frac{i[\sigma_x^1\sigma_x^j,\sigma_z^1]}{2}\right)}\notag\\
&= \exp{(-i\frac{\pi}{4} \sigma_x^1\sigma_x^j)} \exp{(-i\theta \sigma_z^1)}  \exp{(i\frac{\pi}{4} \sigma_x^1\sigma_x^j)}\notag\\
&= \exp\left(i\frac{\pi}{4} \sigma_x^1\sigma_x^j\right) \exp\left(-i(\theta+\pi) \sigma_z^1\right)  \exp\left(i\frac{\pi}{4} \sigma_x^1\sigma_x^j\right).\notag
\end{align}

\subsection{Data for molecular hydrogen}
\label{sec:H2-VQE-fit}
Figure \ref{sfig:H2VQE}.a shows an example of an online VQE run for $R=0.5$~\AA, using the basic Nelder-Mead optimization routine. The experimental data closely follows the theoretical simulation and appears to converge after 15 iterations. In order to maintain the same analysis that was used to accommodate cases in which the optimization routine got stuck, the corresponding energy minimum is extracted by fitting a sinusoidal function to the parameter space explored by all iterations (see Fig~\ref{sfig:H2VQE}.b). The vertical displacement between theory simulation and experiment fits are likely due to noise. The horizontal displacement is likely due to calibration drifts, leading to a different $\alpha$ value having the smallest energy.

\subsection{Decoherence simulation}
\label{sub:H2decoherenceSim}

To understand the effect of various decoherence channels, we performed a simulation of the entire circuit of the two-qubit  H$_2$ experiment using the open source framework OpenFermion \cite{McClean:2017b}. 

We assume that throughout the execution of the entire state preparation circuit, the qubits experience dephasing, e.g. due to magnetic field fluctuations induced by the environment. We model the dephasing via an i.i.d. channel of the form $$\epsilon(\rho) = \epsilon_{i_1} \circ \epsilon_{i_2} (\rho) $$ where
\begin{equation*}
\epsilon_{i}(\rho) = (1 - p_d) \rho + p_d \sigma^{z}_i \rho \sigma^{z}_i,
\end{equation*}
is a Kraus map and $p_d$ is the probability for a single phase flip. In our simulations, we applied the dephasing channel to all the qubits after the application of each gate in the circuit of Figure ~\ref{fig:H2-orb-circ-expval}.a, with probability \mbox{$p_d = 1.0 - \exp(-T_g/T_2)$}, where $T_g$ is the physical time of the gate and $T_2$ is the dephasing time. We employed $T_2=40$~ms, as determined from Ramsey experiments on a single ion. In addition to dephasing, we model the effect of errors in the MS gates using a two-qubit depolarizing channel.
Here, we consider all single- and two-qubit errors with the same probability. For a 2-qubit MS gate, the noise is described by the quantum operation
\begin{align*}
\epsilon_{MS}(\rho) = & (1-p_{MS}) \rho + \frac{p_{MS}}{15} \sum_{i\in\Lambda_a} \sum_{\alpha \in \Lambda_\alpha} \sigma_i^{\alpha} \rho \sigma_i^{\alpha} \\ & + \frac{p_{MS}}{15} \sum_{j_1, j_2\in\Lambda_a} \sum_{\alpha, \beta \in \Lambda_\alpha} \sigma_{j_1}^{\alpha} \sigma_{j_2}^{\beta} \rho \sigma_{j_1}^{\alpha} \sigma_{j_2}^{\beta}, 
\end{align*}
where $p_{MS}$ is the probability of a MS depolarizing error, and $\Lambda_{a}$ and $\Lambda_{\alpha}$ correspond to the set of indexes for the active ions and Pauli matrices, respectively. For the 2-qubit MS gate, there are 15 possible Pauli errors (6 single-qubit and 9 two-qubit), resulting in the prefactor 1/15. The probability $p_{MS}$ is related to the fidelity of the gate as $F = 1 - \frac{14}{15} p_{MS}$ \cite{Bermudez:2017}. The experimental fidelity estimated for the 2-qubit MS gate is 0.99. In the simulation, the two-qubit depolarizing channel is applied with probability $p_{MS}$ after the MS gate.

\begin{figure}[!t] 
   \includegraphics[scale=1.0]{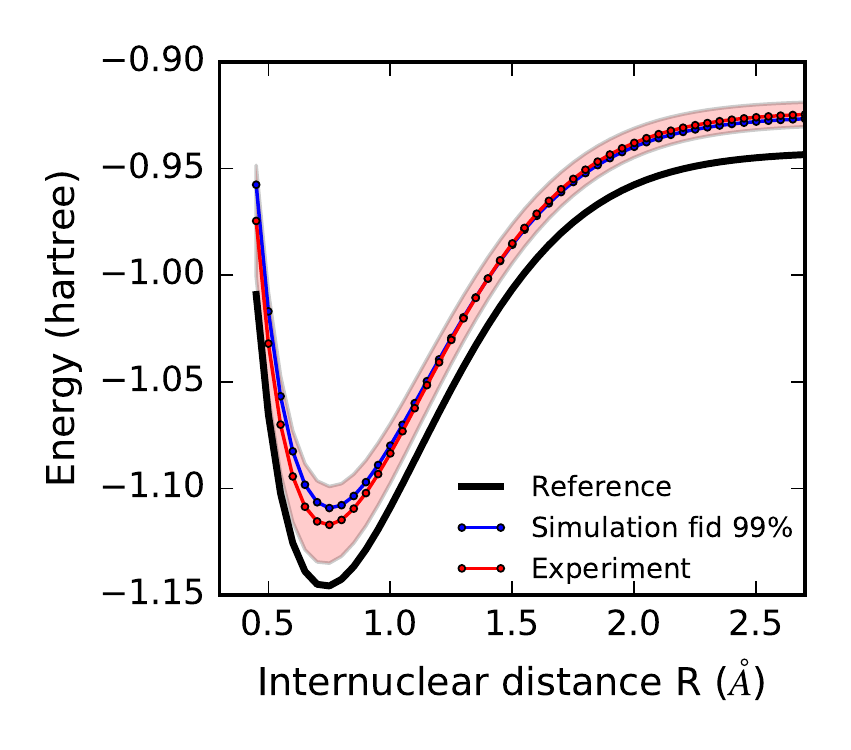} 
   \caption{Simulation of decoherence channels in H$_2$ under the BK mapping. The plot compares the simulated energy curves for different values of the fidelity of the MS gate and the experimental results. The reference curve corresponds to the exact diagonalization of the Hamiltonian. The decoherence channels include one-qubit dephasing acting on all the qubits during the state preparation and two-qubit depolarizing errors in the MS gate.}
\label{sfig:H2decoherence}
\end{figure}

Figures~\ref{fig:energy_errors} and \ref{sfig:H2decoherence} display the results of the simulation for H$_2$ under the BK mapping. Our simulation appears to account for the observed experimental errors along a significant portion of the energy curve. We observe an uneven upshift of the energy values and effectively reduces the estimated well depth, respectively binding energy. While our simulation are close in magnitude to the observed results, we note that other factors such as faulty measurement operators (related to basis rotations and detection fidelity) could also contribute to the discrepancies.

\section{Lithium Hydride molecule}
\subsection{Derivation of the Hamiltonian}
\label{sup:LiH}

We obtained the molecular integrals \eqref{eq:integrals} for LiH at different internuclear separations using a STO-6G
basis set. For this particular application, we used the integrals in the natural orbital basis (NMO). NMOs are obtained by diagonalizing the exact one-electron reduced density matrix (1-RDM) of the system and are ordered by natural orbital occupation numbers (NOONs). It has been shown that NMOs with small NOONs or NOONs close to full-occupancy have a negligible effect in the electron correlation and therefore can be discarded \cite{davidson:1972}. Consequently, approximate NMOs and NOONs, obtained from perturbation theory or truncated configuration interaction (CI) calculations, are usually employed to reduce the computational cost of more involved correlated calculations and for the selection of active spaces \cite{veryazov:2011}.

\begin{figure}[t!]
\begin{center}
\includegraphics[scale=1.0]{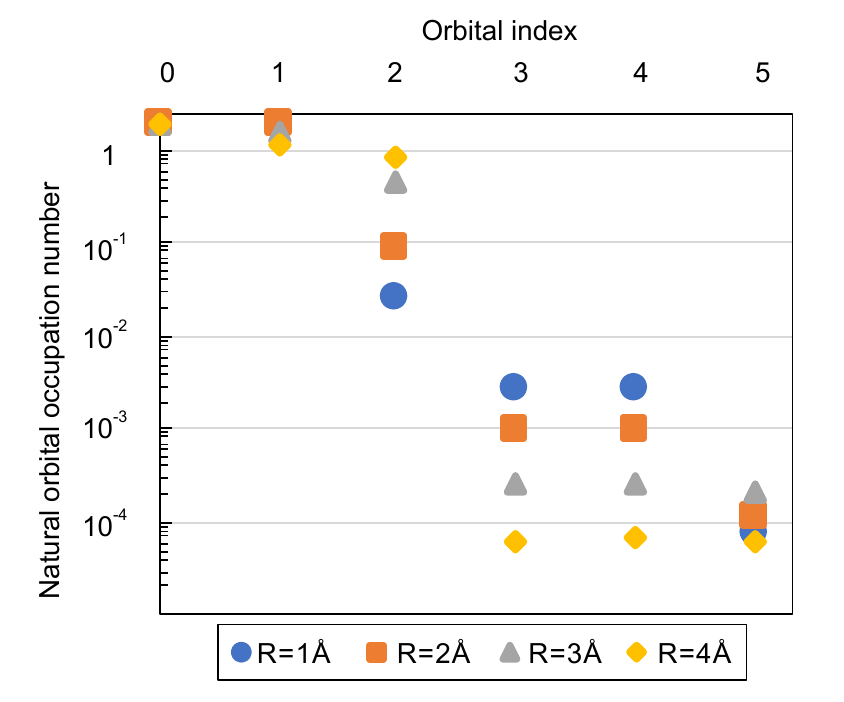}
\caption{Natural orbital occupation numbers at four different internuclear separations for LiH, calculated at the CISD level.}
\label{fig:CISD_amplitudes}
\end{center}
\end{figure}

After the BK transformation, the Hamiltonian for LiH comprises 193 terms with amplitudes larger than $10^{-10}$~Hartree. A VQE simulation using the UCC ansatz truncated to single and double excitations requires
12 qubits and involves 32 single excitation operators and 168 double excitation operators (without imposing spin constrains). To reduce the number of excitation operators in the calculation, we employed the NOONs derived from a configuration interaction calculation with single and double excitations (CISD) in order to select an appropriate active space. Fig.~\ref{fig:CISD_amplitudes} shows the NOONs for the six molecular orbitals of LiH, calculated using CISD for four different internuclear separations. Based on the variations in the NOONs, we establish orbitals 1 to 4 as an appropriate active space.

A reasonable choice of excitation operators in the selected active space would be the singlet double excitations from orbital 1 to orbitals 2, 3 and 4, respectively. Based on the NOONs, we expect the amplitude of the excitation operator from orbital 1 to orbital 2 to be largest. Similarly, we expect excitation operators from orbital 1 to orbitals 3 and 4 to have the same or similar amplitudes. Due to constrains in circuit depth, we consider only excitations from orbital 1 to orbitals 2 and 3, explicitly the operators: $a^{\dagger}_5 a^{\dagger}_4 a_3a_2 - a^{\dagger}_2 a^{\dagger}_3 a_4a_5$ and $a^{\dagger}_7 a^{\dagger}_6 a_3a_2 - a^{\dagger}_2 a^{\dagger}_3 a_6a_7$, where $a^{\dagger}_i (a_i)$ denote the creation (annihilation) operator in the $i$-th spin-orbital. In our notation, spin-orbitals with odd (even) indices correspond to spin-up (spin-down) electrons, with indices starting at 0.
Using the BK mapping, these operators can be expressed as
\begin{align}
a^{\dagger}_5 &a^{\dagger}_4 a_3a_2 - a^{\dagger}_2 a^{\dagger}_3 a_4a_5 \equiv \nonumber \\ 
\frac{i}{8} (&\sigma_2^X \sigma_4^Y + \sigma_1^Z\sigma_2^X\sigma_3^Z\sigma_4^Y -\sigma_2^Y\sigma_4^X  - \sigma_1^Z\sigma_2^Y\sigma_3^Z\sigma_4^X 
- \sigma_2^Y\sigma_4^X\sigma_5^Z  \notag\\&-\sigma_1^Z\sigma_2^Y\sigma_3^Z\sigma_4^X\sigma_5^Z \nonumber +\sigma_2^X\sigma_4^Y\sigma_5^Z  +\sigma_1^Z\sigma_2^X\sigma_3^Z\sigma_4^Y\sigma_5^Z)\nonumber
\end{align}

\noindent and 

\begin{align}
a^{\dagger}_7 &a^{\dagger}_6 a_3a_2 - a^{\dagger}_2 a^{\dagger}_3 a_6a_7 \equiv \nonumber\\
\frac{i}{8} (&\sigma_2^X \sigma_6^Y + \sigma_1^Z\sigma_2^X\sigma_3^Z\sigma_6^Y -\sigma_2^Y\sigma_6^X - \sigma_1^Z\sigma_2^Y\sigma_3^Z\sigma_6^X\notag\\
 &-\sigma_2^Y\sigma_3^Z\sigma_5^Z\sigma_6^X\sigma_7^Z 
- \sigma_1^Z\sigma_2^Y\sigma_5^Z\sigma_6^X\sigma_7^Z \nonumber \\
&+ \sigma_2^X\sigma_3^Z\sigma_5^Z\sigma_6^Y\sigma_7^Z   
+\sigma_1^Z\sigma_2^X\sigma_5^Z\sigma_6^Y\sigma_7^Z)\nonumber
\end{align}

\begin{figure}[t!]
 \centering
 \includegraphics[scale=1]{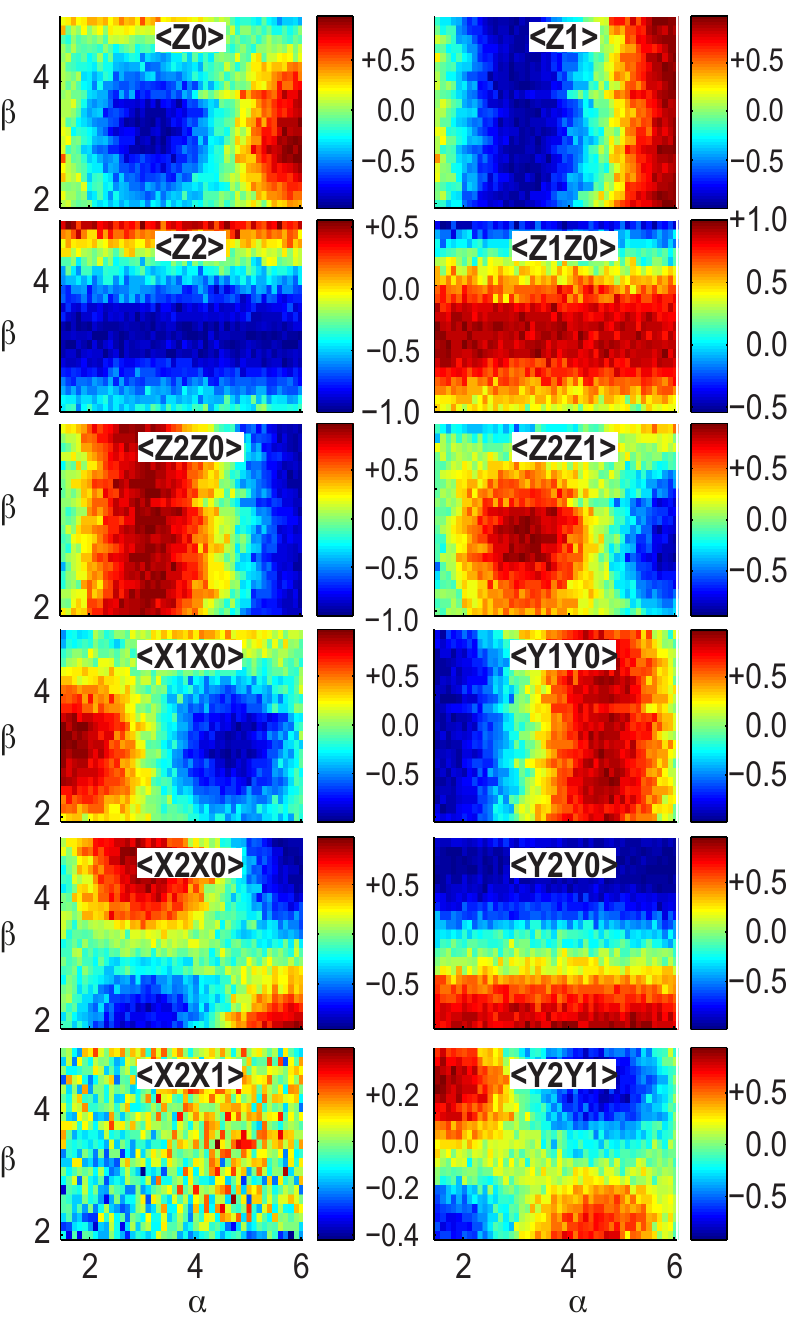}
 \caption{Experimentally measured expectation values of the Pauli operators $H_\ell$ of Hamiltonian \ref{eq:LiH_Hamiltonian} as a function of circuit parameters $\alpha$ and $\beta$.}
 \label{fig:LiHscan_expval}
\end{figure}

The initial state of the simulation, corresponding to the Hartree-Fock wavefunction, is the state $\ket{000000000101}$, which simplifies to $\ket{000001}$ in the active space required for the selected excitation operators. A full-simulation of LiH with the two double excitation operators listed above would require at most 32 MS gates for a single Trotter step if all the subterms were going to be implemented. To make the simulation affordable in the current device we approximated each of the operators using only the first subterm, corresponding to $\sigma_2^X\sigma_4^Y$ and $\sigma_2^X\sigma_6^Y$, respectively. The absolute error in the total energy introduced by this approximation is smaller than chemical accuracy within the basis set used, when compared to the full configuration interaction (FCI) reference solution. The two selected subterms can be implemented using MS gates and single qubit rotations as follows

\begin{align}
\exp[-i\alpha \sigma_2^X\sigma_4^Y] \equiv U^{\{2,4\}}_{MS}(\frac{\pi}{2},0) R_z(\alpha,2)U^{\{2,4\}}_{MS}(-\frac{\pi}{2},0) \notag\\
\exp[-i\beta \sigma_2^X\sigma_6^Y] \equiv U^{\{2,6\}}_{MS}(\frac{\pi}{2},0) R_z(\beta,2)U^{\{2,6\}}_{MS}(-\frac{\pi}{2},0)\notag
\end{align}

where $R_z$ represents a rotation around the $Z$-axis and $U^S_{MS}$ is an MS gate acting on the set of qubits $S$. As the entangling operations involve only qubits $2$, $4$ and $6$, we can efficiently construct
an effective Hamiltonian involving only operations on these qubits. The corresponding 3-qubit Hamiltonian has the form

\begin{align}
H =&c_0 I+ c_1 Z_0+ c_2 Z_1+ c_3 Z_2+c_4 Z_1Z_0+c_5 Z_2Z_0\nonumber \\
&+ c_6 Z_2Z_1+c_7 X_1X_0+c_8Y_1Y_0+c_9 X_2X_0\nonumber \\
&+ c_{10} Y_2Y_0+c_{11} X_2X_1+ c_{12} Y_2Y_1.
\label{eq:LiH_Hamiltonian}
\end{align}

The applied pulse sequence to realize the Hamiltonian in Eq.~(\ref{eq:LiH_Hamiltonian}) is shown in Figure \ref{fig:LiHconcept}.b acting on initial state $\ket{111}$. 

\begin{figure}[!t]
 \centering
 \includegraphics[scale=1]{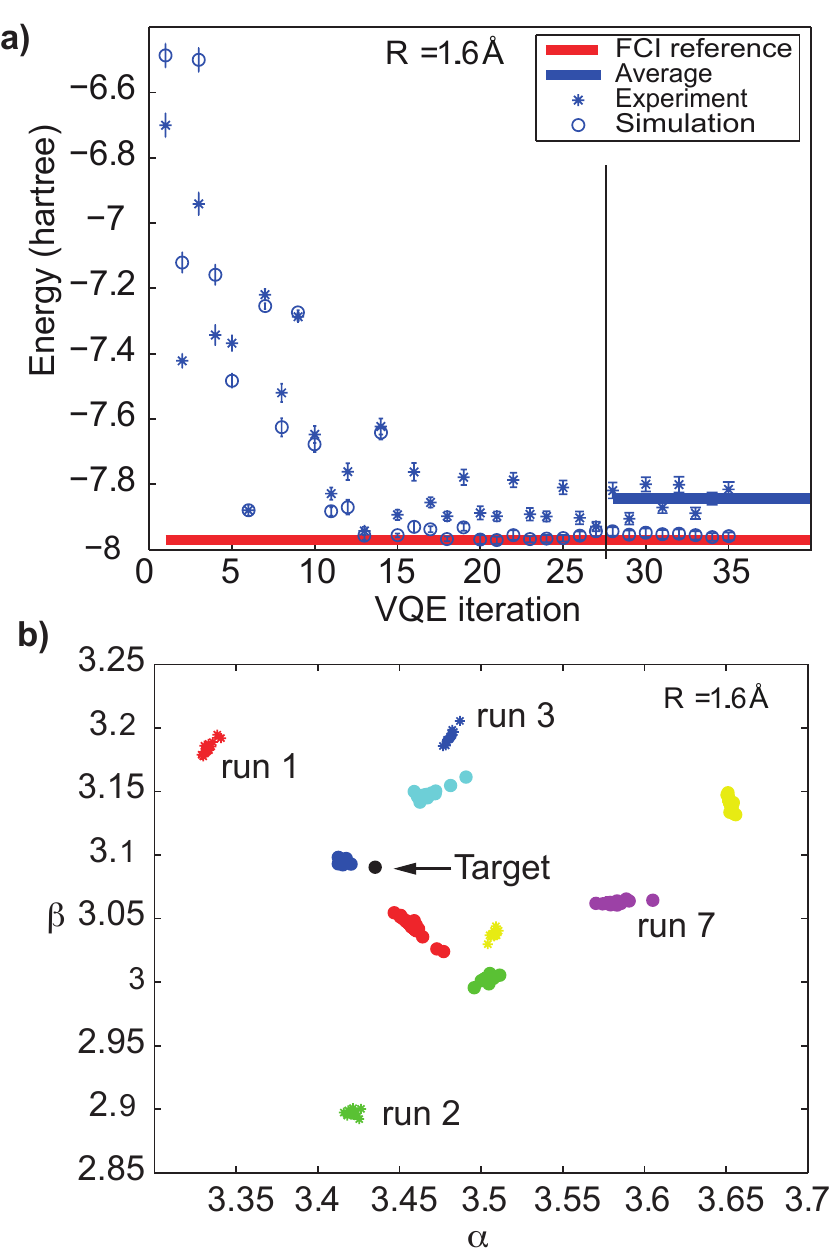}
 \caption{\textbf{a)} VQE run for LiH under the basic Nelder-Mead direct search algorithm,   for nuclear separation $R=1.6$~\AA. The graph shows Energy $\expval{H(R)}_{\alpha,\beta}$ vs. iteration number. Stars represent experimental data with errorbars derived from quantum projection noise. Circles represent a theoretical simulation of the experiment. The red line indicates the theoretical FCI value, the blue line shows the average over the last experimental iterations, between the black line and last point. 
 \textbf{b)} Last 20 steps of 10 independent simulations of the VQE experiment with quantum projection noise, using the basic Nelder-Mead direct search algorithm at $R=1.6$~\AA. The black dot indicates the optimum parameter combination for the global minimum location, predicted by FCI theory. The colored clusters of symbols represent different repetitions of the same simulation that all started at the same initial values of $\alpha$ and $\beta$.}
 \label{LiHVQEsim}
\end{figure}

\subsection{Parameter scans for LiH}
\label{sec:LiH_scan}

The 2D LiH energy landscape $\expval{H(R)}_{\alpha,\beta}$, as shown in the inset of  Fig.~\ref{fig:LiH}.a), is obtained in the following way:
We apply the pulse sequence, shown in Fig.~\ref{fig:LiHconcept}.c) to the initial state $\ket{\varphi(0)} = \left | 111 \right \rangle$. The MS gate fidelity for three ions reaches $97(3) \%$ and the experiment is performed with 100 repetitions.
The energy depends on two parameters $\alpha,\beta$, both corresponding to a rotation angle on $\sigma_z$, applied on one of three ions. 
These angles are scanned in the range of $\alpha\in \left [1.5,6  \right ],\beta\in \left [2,5  \right ]$ with resolutions of $0.1$ and $0.15$ respectively. 
The measured expectation values are shown in Fig.~\ref{fig:LiHscan_expval}. 
The 2D energy landscapes $\expval{H(R)}_{\alpha,\beta}$, as shown in Fig.~\ref{fig:LiH}.a), are finally calculated by combining the measured expectation values according to Eq.~(\ref{eq:LiH_Hamiltonian}) and adding the contribution from the Coulomb interaction of the nuclei $\expval{H_{nuc}(R)}$ (Eq.~\ref{eq:energy}). 

The final molecular energy curve $\expval{H(R)}$ for LiH, as shown in Fig.~\ref{fig:LiH}.b, is derived by fitting two-dimensional functions on the 2D energy landscapes and extracting the minimum value. 

\subsection{VQE runs for LiH}
\label{sup:VQELiH}

In our first implementation of the VQE we employed the Nelder-Mead optimization routine in the same way as for molecular hydrogen. In the experimental results shown in Figure \ref{LiHVQEsim}.a it appears that the optimization algorithm becomes trapped in a local minimum, preventing it from converging any further. 

\begin{figure}[!h]
 \centering
 \includegraphics[scale=1]{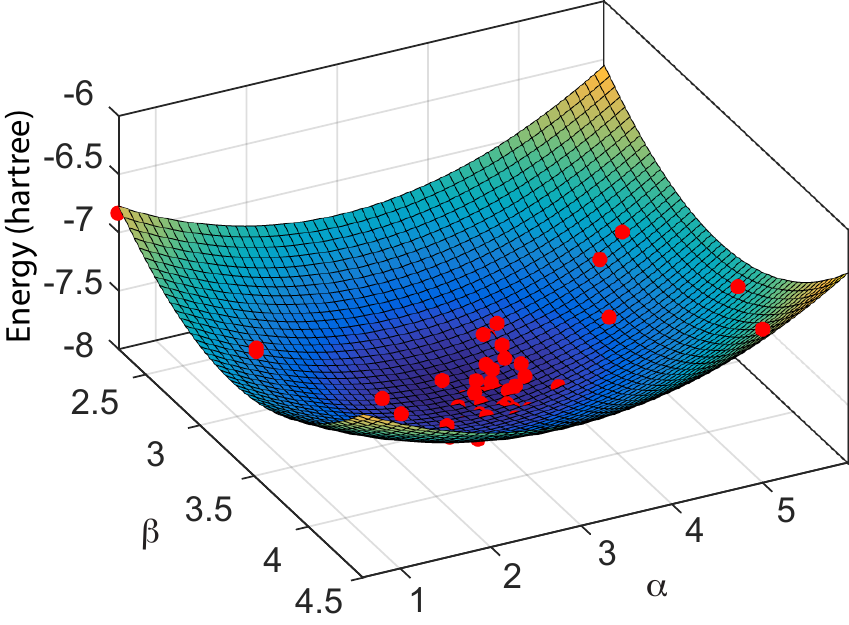}
 \caption{Fitting of the two-dimensional quadratic function $E(\alpha,\beta) = m+(c \alpha-a)^2+(d  \beta-b)^2$ to the VQE points (red). 
 }
 \label{fig:LiH_3D}
\end{figure}

To investigate this phenomenon, we simulate the experiment, including quantum projection noise, and run the a full VQE simulation 10 times at a fixed separation R using the same values for initial guess of parameters $\alpha$ and $\beta$. The simulation results are shown in Figure~\ref{LiHVQEsim}.b. For better visibility, we plot only the last 20 points for each run, after which the algorithm converges to a ``stable" position. The black dot represents the optimal combination of parameters, predicted by the theory. 
One finds, that each of the 10 simulated runs is trapped in a different local minimum, leading to a deviation from the ideal energy values. We conclude that the basic Nelder-Mead algorithm is not suitable for this optimization problem.
We therefore instead employ a hybrid of the Nelder-Mead direct search algorithm and simulated annealing theory \cite{Vugrin:2005} to obtain a better estimate of the energy $\expval{H(R)}_{\alpha,\beta}$ in the noisy environment.
We now fit a quadratic function to the 2D landscape and extract the minimum value as $\expval{H(R)}$. Fig.~\ref{fig:LiH_3D} shows an example of fitting the 2D quadratic function  $E(\alpha,\beta) = m+(c\cdot \alpha-a)^2+(d\cdot \beta-b)^2$, with the minimum energy value $m$, to a subset of the VQE iterations, taken to include those within 4 standard deviations from the median. 
We perform VQE runs for two different internuclear separations $R = 1.6, R = 2.75$, with 3 ion MS gate fidelities of $98(5) \%$. The results are shown in Fig.~\ref{fig:LiH}.b, together with the parameter scan discussed in the previous section.

\end{document}